\documentclass[11pt,graphicx]{article}
\usepackage{graphicx}
\usepackage{amsmath}
\usepackage{amsfonts}
\usepackage{amssymb}
\usepackage{dsfont}
\usepackage{dcolumn}
\usepackage{bm}

\def\beq{\begin{equation}}
\def\be{\begin{equation}}
\def\beqn{\begin{eqnarray}}
\def\ee{\end{equation}}
\def\eeq{\end{equation}}
\def\eeqn{\end{eqnarray}}

\def \cha{\widetilde{\chi}^{\pm}_1}

\def \na{\widetilde{\chi}^{0}_1}
\def \nb{\widetilde{\chi}^{0}_2}

\def \g{\widetilde{g}}

\def \ta{\widetilde{t}_1}

\def \sta{\widetilde{\tau}_1}
\def \stb{\widetilde{\tau}_2}

\def \slr{\widetilde{l}_R}

\def \snl{\widetilde{\nu}_{\tau}}
\def \snm{\widetilde{\nu}_{\mu}}

\def \hc{H^{\pm}}

\begin{document}
\baselineskip 18pt

\begin{center}{\LARGE\bf
Light Higgses at the  Tevatron and at the
LHC and Observable Dark Matter in SUGRA and D-branes
 }\\
\vskip.25in {Daniel Feldman\footnote{E-mail: feldman.da@neu.edu},
Zuowei Liu\footnote{E-mail: liu.zu@neu.edu} and Pran
Nath\footnote{E-mail: nath@lepton.neu.edu}  }\\
{\it
 Department of Physics, Northeastern University, Boston, MA 02115, USA\\
}
\end{center}


\begin{abstract}
Sparticle landscapes in mSUGRA, in SUGRA models with
nonuniversalities (NUSUGRA), and in  D-brane models are analyzed.
The analysis exhibits the existence of Higgs Mass Patterns (HPs)
(for $\mu>0$) where the CP odd Higgs could be the next heavier
particle beyond the LSP and sometimes even lighter than the LSP. It
is shown that the Higgs production cross sections from the  HPs are
typically  the largest enhancing the  prospects for
 their  detection at the  LHC.
Indeed it is seen that the recent Higgs  production limits from
CDF/D\O\ are beginning to put constraints on the HPs. It is also
seen that the $B_s\to \mu^+\mu^-$ limits  constrain the HPs more
stringently. Predictions of the Higgs  production cross sections for
these patterns at the LHC are  made.   We compute the
neutralino-proton cross  sections  $\sigma (\tilde\chi_1^0 p)$  for dark
matter experiments and show that the largest $\sigma (\tilde\chi_1^0 p)$ also
arise from the HPs and further that  the HPs and some of the other
patterns are beginning to be constrained by the most recent data
from CDMS and from Xenon10 experiments. Finally,  it is shown that
the prospects are bright for the discovery of dark matter with
$\sigma(\tilde\chi_1^0 p)$ in the range $10^{-44\pm .5}$cm$^2$ due to a
``Wall" consisting of a copious number of parameter points in the
Chargino Patterns (CPs) where the chargino is the NLSP.  The Wall,
which appears in all models considered (mSUGRA, NUSUGRA and D-branes)
 and runs up to about a TeV in LSP mass,   significantly enhances the chances for the
observation of dark matter
by SuperCDMS,  ZEPLIN-MAX, or LUX  experiments  which are expected to
achieve  a sensitivity of $10^{-45}$ cm$^2$ or more.
\end{abstract}

Recently a new approach  for the search for sparticles at colliders
was  given in the framework of sparticle landscapes
\cite{Feldman:2007zn}.  In this work it is shown that while in the
MSSM,  which has 32 sparticles, the sparticle masses can generate as
many $10^{28}$ mass hierarchies,  the number of these mass
hierarchies decreases  enormously  in well motivated models such as
gravity mediated breaking models \cite{chams,barbi}.
 It  is further shown that
if one limits one self to the first four sparticles aside from the
lightest Higgs boson, then the number of such possibilities reduces
even further.  Specifically within the  minimal supergravity grand unified
model \cite{chams}, mSUGRA, which has a parameter space defined
by \cite{chams,hlw}
 $m_0,m_{1/2}, A_0, \tan\beta$ and the sign of the Higgs mixing parameter
 $\mu$,
one finds that the number of such patterns reduces to 16  for
$\mu>0$, and these patterns are labeled mSP1-mSP16
\cite{Feldman:2007zn}. These patterns are further classified by the
next  to the lightest sparticles beyond the LSP which are  found to
be the chargino for $ \rm mSP(1-4)$, the stau  for $\rm mSP(5-9)$,
the stop for $\rm mSP(11-13)$, and the Higgs $A/H$, where $A$ is the
CP odd Higgs in the MSSM and $H$ is the heavier CP even Higgs,  for
$ \rm mSP(14-16)$. Thus the patterns are labeled the Chargino
Pattern, the Higgs Pattern etc.
  Most of these patterns appear to have  escaped attention
in previous studies because the parameter searches were based on
 restricted   regions of the parameter space. In our analysis we have carried out an exhaustive search under
naturalness assumptions in exploring the sparticle landscape and the
residual parameter space which satisfies the radiative electroweak
symmetry breaking  (REWSB), the dark matter relic density
constraints and the collider constraints from flavor changing
neutral currents and sparticle mass limits. The analysis exhibits a
much larger set of patterns than previously seen.

 In the analysis presented here we consider  a larger  class of models than discussed in \cite{Feldman:2007zn}.
 Specifically we consider  mSUGRA models  (for recent
works on mSUGRA  see, e.g., \cite{Djouadi:2006be}) with both signs
of $\mu$
   as well as  SUGRA models with nonuniversalities (NUSUGRA),
  and  D-brane models.  The  focus of our work will be Higgs Patterns
 which we   collectively call  HPs.
 It will be  shown
that typically the HPs  lead to the largest production cross
sections for the
CP even and CP odd  Higgs  at  the Tevatron and at
the LHC. Further, they also lead to an LSP which  has  a very
substantial Higgsino component.  It is also shown that the HPs lead
to the largest  branching ratio for $B_s\to \mu^+\mu^-$. Finally, we
show that the  largest  spin independent neutralino-proton cross
section in dark matter experiments also arises from the HPs  and the
most recent results from  the dark matter experiment are beginning
to constrain the HPs,  and more generally the dark matter
experiments  can also  serve as a discriminator amongst
sparticle  mass  patterns in the landscape.

We begin by discussing the details of the analysis.
For the relic density  of the neutralino LSP we impose the WMAP3
constraints \cite{Spergel:2006hy},
 $0.0855<\Omega_{\na} h^2<0.1189~(2\sigma)$.  As is well known the
 experimental limits on  the  FCNC process $b\to s\gamma$ impose severe constraints
 and we use here the  constraints from
the Heavy Flavor Averaging Group (HFAG) \cite{hfag} along with the
BABAR, Belle and CLEO
   experimental results: ${\mathcal Br}(b\rightarrow s\gamma) =(355\pm 24^{+9}_{-10}\pm 3) \times 10^{-6}$.
A new estimate of ${\mathcal Br}(\bar B \to X_s \gamma)$ at
$O(\alpha^2_s)$ gives \cite{Misiak:2006zs} ${\mathcal
Br}(b\rightarrow s\gamma) =(3.15\pm 0.23) \times 10^{-4}$  which
moves the previous SM mean value of $3.6\times 10^{-4}$ a bit lower. In
the analysis we use a $3.5\sigma$ error corridor around the HFAG
value. The total ${\mathcal Br}(\bar B \to X_s \gamma)$ including
the sum of SM and SUSY contributions (for the update  on  SUSY
contributions see \cite{susybsgamma}) are constrained by  this
corridor.
    The process $B_s\to \mu^+\mu^-$ can become significant for large
   $\tan\beta$ since
  the decay  has a  $\tan^6\beta$
  dependence and thus large $\tan\beta$ could be constrained by the
  current limit which is
${\mathcal Br}( B_s \to \mu^{+}\mu^{-})$ $< 1.5 \times10^{-7}$ (90\% CL), $ 2.0 \times
10^{-7}$  (95\% CL) \cite{Abulencia:2005pw}.
 We note that more recently the CDF and
D\O\  have  given limits which are about a factor of 10 more
sensitive. We have included these preliminary \cite{bsmumu07}
results in this analysis. Additionally, we also impose the current
lower limits on the lightest  CP even  Higgs boson. For the Standard
Model like Higgs boson this limit is
  $\approx$ 114.4 {~\rm GeV} \cite{smhiggs}, while a limit of 108.2 {\rm GeV} at 95\% CL
  is set on the production of an invisibly decaying Standard Model like Higgs by
  OPAL \cite{smhiggs}.
   For the MSSM we take the constraint to be $m_h> 100 ~{\rm GeV}$.
 We take the other sparticle mass
  constraints to be $m_{\cha}>104.5 ~{\rm GeV}$ \cite{lepcharg} for the lighter
 chargino,  $m_{\ta}>101.5 ~{\rm GeV}$,  $m_{\sta}>98.8 ~{\rm  GeV}$ for the
 lighter   stop and the stau \cite{Djouadi:2006be}.
The mSUGRA analysis is based on a large Monte Carlo scan of the parameter space with
the soft parameters in the range
 $0<m_0<4000 ~{\rm GeV},$ ~$0<m_{1/2}<2000~{\rm GeV}$,  ~$|A_0/m_0|<10, ~1<\tan\beta<60$
 and  both signs of $\mu$ are analyzed.
 In our analysis we  use   MicrOMEGAs version 2.0.7 \cite{MICRO}  which
 includes  the SuSpect 2.34 package \cite{SUSPECT} for the analysis of
 sparticle masses,   with $m^{\overline{\rm MS}}_b(m_b)= 4.23~{\rm GeV}$, ~$m_t(\rm pole) = 170.9 ~{\rm GeV}$,
 requiring REWSB at the SUSY scale.  We have cross checked
 with other codes
 \cite{ISAJET,SPHENO,SOFTSUSY,Baer:2005pv,Belanger:2005jk,Allanach:2004rh,Allanach:2003jw}
  and find  agreement up to $\sim O$(10\%).

 In the analysis a scan of $2\times 10^6$ models  with Monte Carlo simulation
 was used  for mSUGRA with $\mu>0$ and a scan of $1\times 10^6$ models for $\mu<0$.
 Twenty two 4-sparticle patterns labeled mSP1-mSP22 survive the constraints from
the radiative  electroweak  symmetry breaking, from the relic
density constraint, and other collider constraints.  mSP1-mSP16
which appear for $\mu>0$ are  defined  in  \cite{Feldman:2007zn}.
For $\mu<0$ all of the patterns in $\mu>0$ case appear except for
the cases mSP10, mSP14-mSP16.  However,  new patterns mSP17-mSP22
appear for $\mu<0$ and are given below
\begin{equation}
\begin{array}{ll}
{\rm mSP17:}  ~\na < \sta < \nb  <\cha ~;
&{\rm mSP18:}  ~\na < \sta < \slr <\ta ~;\\
{\rm mSP19:}  ~\na < \sta < \ta  <\cha ~;
&{\rm mSP20:}  ~\na < \ta  < \nb  <\cha ~;\\
{\rm mSP21:}  ~\na < \ta  < \sta <\nb ~;
&{\rm mSP22:}  ~\na < \nb  < \cha <\g ~.
\end{array}
\end{equation}
A majority of the patterns discussed in \cite{Feldman:2007zn} and this  analysis do not appear in the
Snowmass  Benchmarks \cite{Allanach:2002nj}, and in the
PostWMAP  Benchmarks \cite{Battaglia:2003ab}. Since the HPs are a focus of this analysis, we exhibit these
below
\begin{equation}
\begin{array}{cc}
(i)   ~{\rm mSP14:}   ~\na  <  A,H  < \hc ~;
&(ii) ~{\rm mSP15:}   ~\na  <  A,H  < \cha ~;\\
(iii) ~{\rm mSP16:}   ~\na  <  A,H  < \sta ~;
&(iv) ~{\rm NUSP12:}  ~\na  <  A,H  < \g   ~,
\label{hps}
\end{array}
\end{equation}
where $A,H$ indicates that the two Higgses $A$ and $H$ may sometimes exchange positions in the sparticle
mass spectra\footnote{In fact there are
cases where all the Higgses $h, H, A, H^{\pm}$ lie below $\na$.}.
\begin{table}[h]
    \begin{center}
    \scriptsize{
\begin{tabular}{|c|c|c|c|c|c|c|c|}
                                                                    \hline  \hline \hline
    &   $m_0$   &   $m_{1/2}$   &   $A_{0}$ &   $\tan {\beta}$  &     NUH &   NUq3 &   NUG \\
{\bf  \rm HPs}   &   (GeV)   &   (GeV)   &   (GeV)   &   $  $   &   $(\delta_{H_u},\delta_{H_d})$   &   $(\delta_{q3},\delta_{tbR})$ &
$(\delta_{M_2},\delta_{M_3})$   \\ \hline  \hline
 ${\bf mSP14 }$ &   1036    &   562 &   500 &   53.5    &      (0,0)   &   (0,0)   &   (0,0)   \\ \hline
 ${\bf mSP14 }$ &   759 &   511 &   2315   &   31.0    &      (0.256,-0.499)   &   (0,0)   &   (0,0)   \\ \hline
 ${\bf mSP14 }$ &   1223 &   1200 &   -111   &   27.4   &     (0.557,-0.736)   &   (0,0)   &   (0,0)   \\ \hline
 ${\bf mSP14 }$ &   740 &   620 &   840 &   53.1    &      (0,0)   &   (-0.553,-0.249) &   (0,0)   \\ \hline
 ${\bf mSP14 }$ &   1201    &   332 &   -731    &   55.0    &      (0,0)   &   (0,0)   &   (0.383,0.275)   \\ \hline \hline
 ${\bf mSP15 }$ &   1113    &   758 &   1097    &   51.6    &      (0,0)   &   (0,0)   &   (0,0)   \\ \hline
 ${\bf mSP15 }$ &   900 &   519 &   1481    &   54.8    &      (0,0)   &   (0,0)   &   (-0.352,-0.262) \\ \hline
 ${\bf mSP15 }$ &   1389    &   551 &   -167    &   59.2    &      (0,0)   &   (-0.041,0.916)  &   (0,0)   \\ \hline \hline
 ${\bf mSP16 }$ &   525 &   450 &   641 &   56.0    &      (0,0)   &   (0,0)   &   (0,0)   \\ \hline
 ${\bf mSP16 }$ &   282 &   464 &   67  &   43.2    &      (0.912,-0.529)  &   (0,0)   &   (0,0)   \\ \hline
${\bf NUSP12}$  &   2413    &   454 &   -2490   &   48.0    & (0,0)
& (0,0)   &   (-0.285,-0.848) \\ \hline
                                                                    \hline \hline
                                                                        \end{tabular}
}
\caption[]{ Benchmarks for HPs for $\mu>0$
 in mSUGRA and in NUSUGRA. The 2nd and the 3rd mSP14 pattern show that the HPs can
 emerge  for moderate  values of $\tan \beta$. The Benchmarks are computed with SuSpect
2.34 .
 }
 \label{b5}
    \end{center}
 \end{table}
 The cases  (i)-(iii) in Eq.(\ref{hps})
arise  for $\mu>0$ and not for $\mu<0$, and the case  (iv) in Eq.(\ref{hps}) arises
in an isolated region of the parameter space  for $\mu>0$  in  the NUSUGRA case discussed later.  The sign of
$\mu$ is   very relevant in the analysis not only because  the HPs
for mSUGRA case arise only for $\mu>0$, but also because of the recent results
from the  $g_{\mu}-2$ experiment.
 As is well known the supersymmetric
electroweak corrections to $g_{\mu}-2$
can be  as large or even larger than the Standard Model
correction \cite{yuan}. Further, for  large $\tan\beta$ the sign of  the supersymmetric
correction to $g_{\mu}-2$ is correlated with the sign of  $\mu$.
The  current data \cite{Bennett:2004pv,Hagiwara:2006jt}
  on $g_{\mu}-2$
favors  $\mu>0$ and thus it is of relevance to discuss the possible
physics that emerges if indeed one of these patterns is the one that
may be  realized in nature. Some benchmarks for the HPs are given in
Table (\ref{b5}).
\begin{figure}[t]
\begin{center}
\includegraphics[width=7.5cm,height=6.5cm]{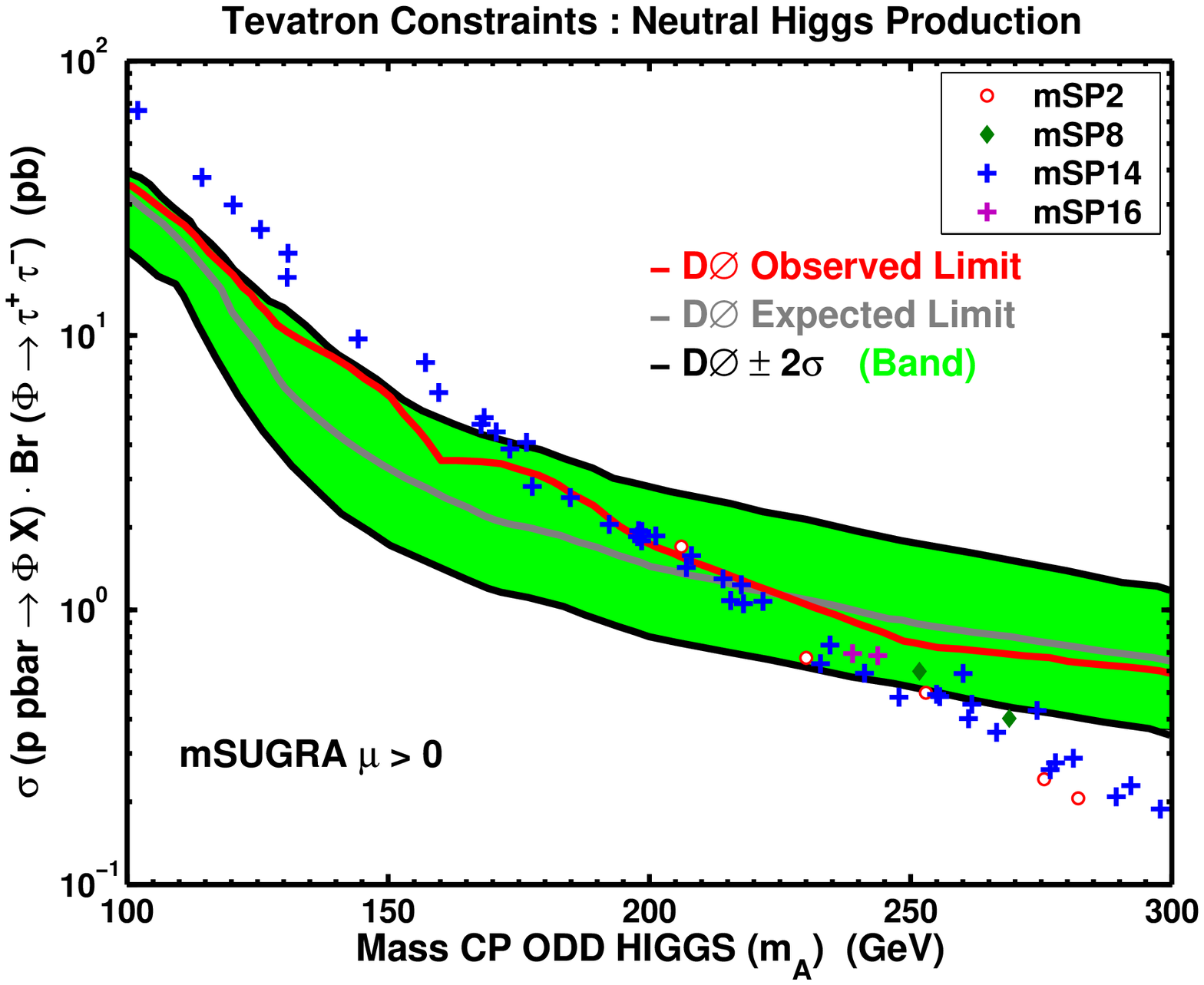}
\includegraphics[width=7.5cm,height=6.5cm]{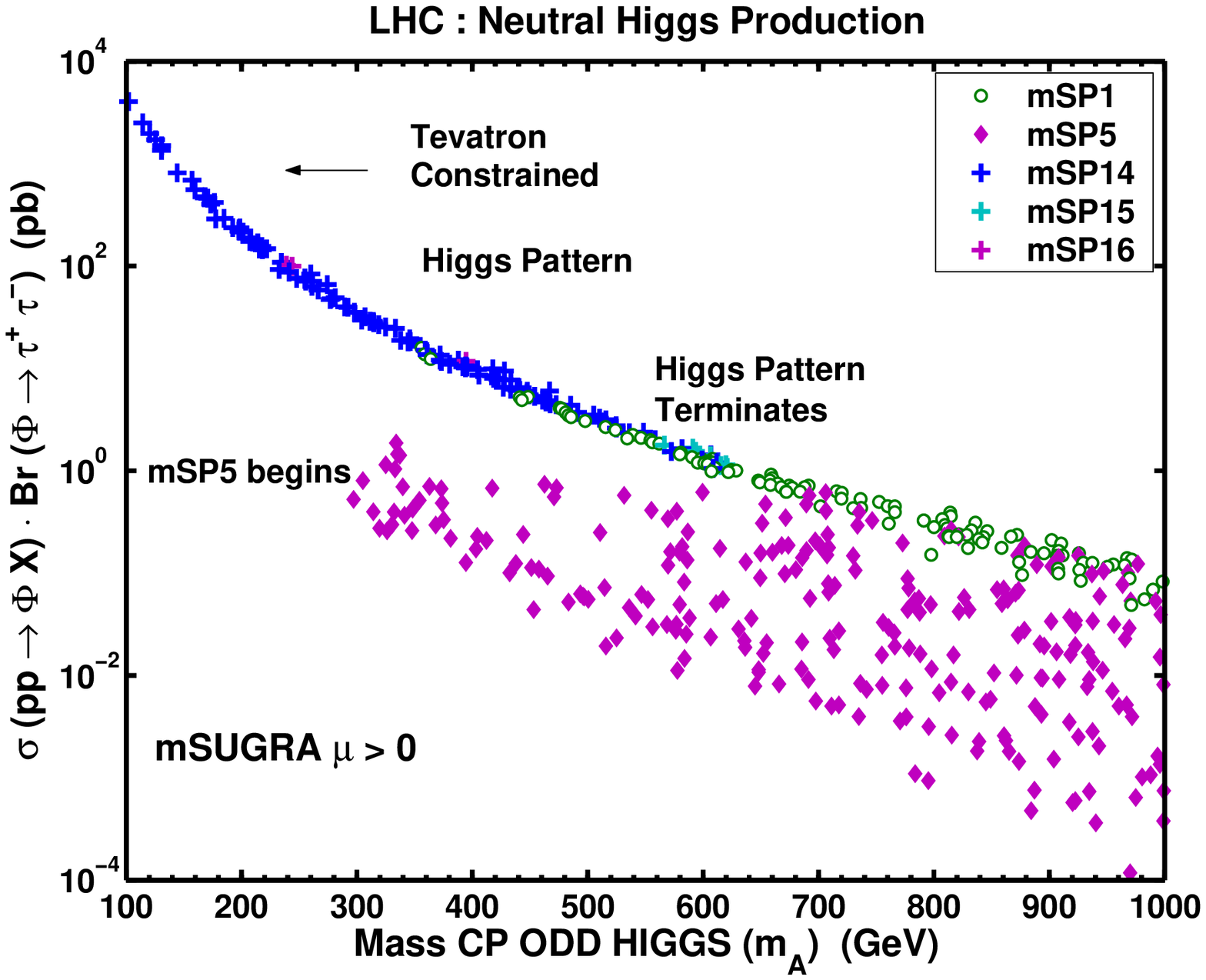}
\caption[]{(Color Online) Left panel: Predictions for $
[\sigma(p\bar p \to \Phi){\rm BR}(\Phi \to 2\tau)]$
 in mSUGRA as a function of the CP odd Higgs mass $m_A$ for
the HPs at the Tevatron with CM energy of $\sqrt s =1.96$ TeV. The
limits from D\O\ are indicated \cite{Abazov:2006ih}.  Right panel:
Predictions for
 $[\sigma(p p \to \Phi){\rm BR}(\Phi \to 2\tau)]$
in mSUGRA as a function of $m_A$ at
the LHC  with CM energy of $\sqrt s =14$ TeV for the HPs, the
chargino pattern mSP1 and the stau pattern mSP5. The HPs are seen
to give the largest cross sections.}
 \label{higgstevlhc}
  \end{center}
\end{figure}

\begin{figure}[t]
\begin{center}
\includegraphics[width=7.5cm,height=6.5cm]{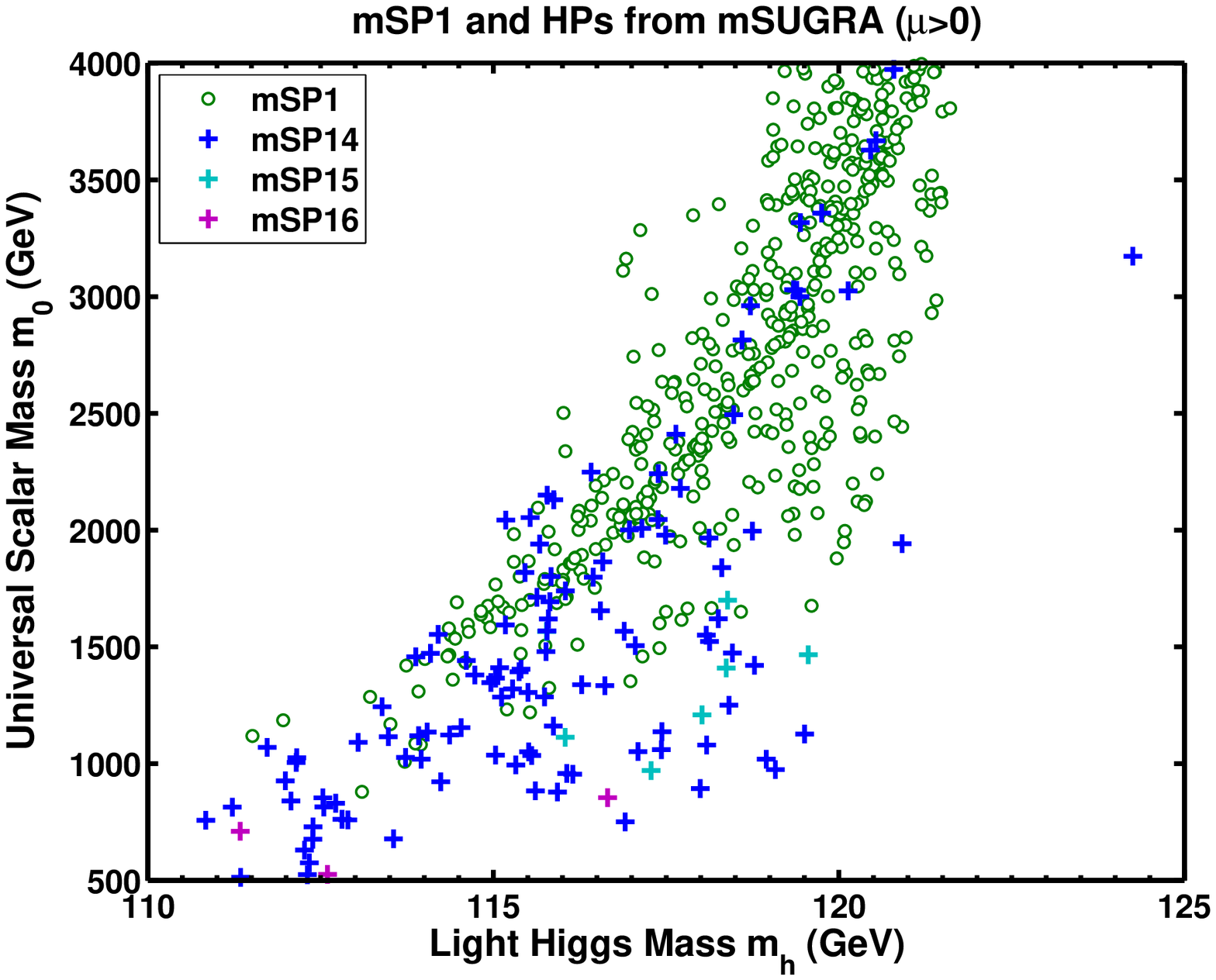}
\includegraphics[width=7.5cm,height=6.5cm]{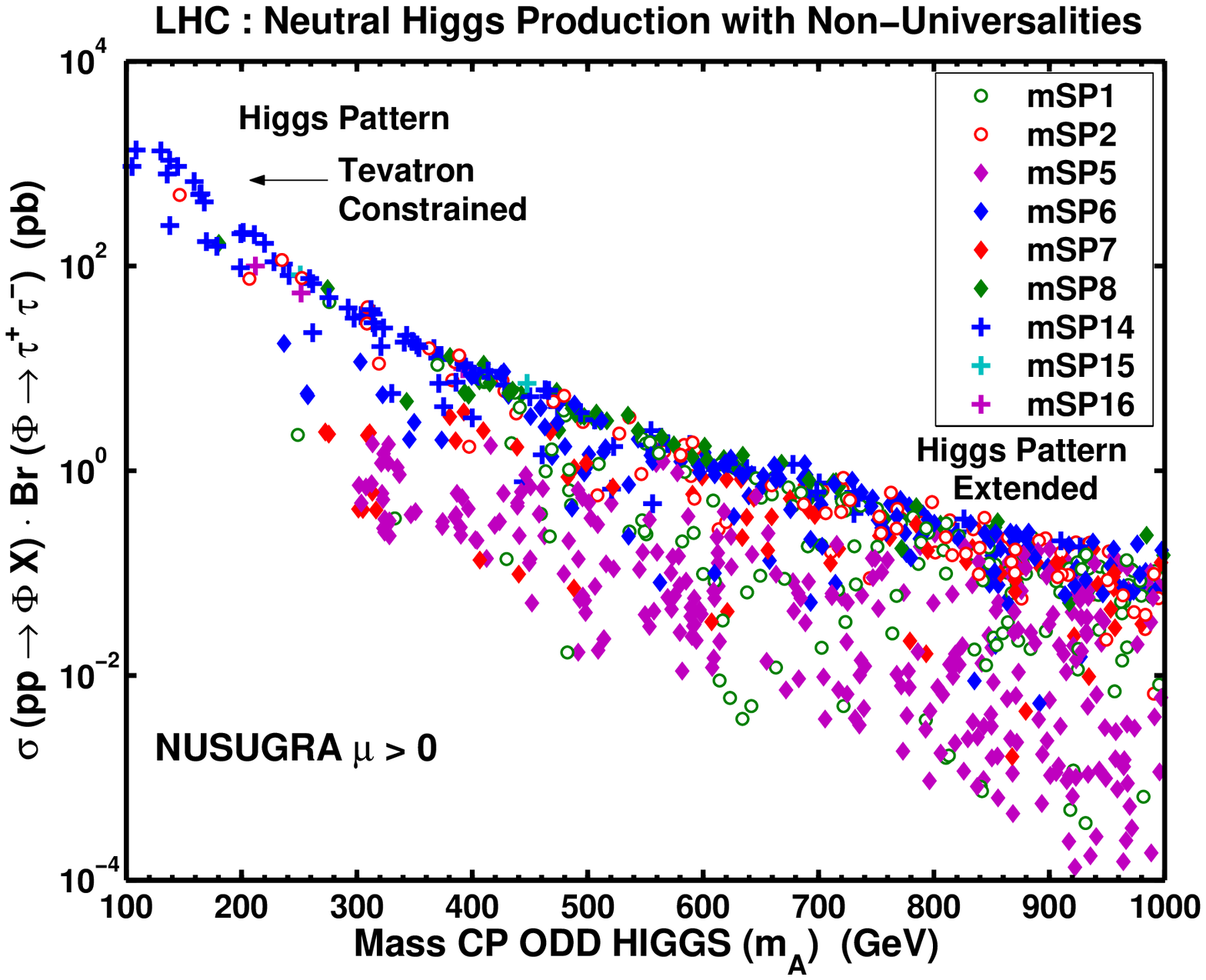}
\caption[]{(Color Online) Left panel: mSP1 and HPs are plotted in
the $m_0$-$m_h$ plane in mSUGRA $\mu > 0$. Right panel: Predictions
for
$  [\sigma(p p \to \Phi){\rm BR}(\Phi \to 2\tau)]$
 in NUSUGRA (NUH,NUG,NUq3) as a function of CP odd Higgs
mass at the LHC showing that the HPs extend beyond 600 ${\rm GeV}$
with non-universalities (to be compared with the analysis of
Fig.(\ref{higgstevlhc}) under the same naturalness assumptions).}
 \label{mandNUcross}
  \end{center}
\end{figure}

{\it
Higgs cross sections at the Tevatron and at the LHC}:
 The lightness of $A$ (and also of $H$ and $H^{\pm}$)
 in the Higgs Patterns   implies that the Higgs  production cross sections can be large
 (for some of the previous analyses where light Higgses appear
 see \cite{Kane:2003iq,Carena:2006dg,Ellis:2007ss}).
 Quite interestingly the recent Tevatron data is beginning to constrain
 the HPs. This is exhibited  in the left panel of
  Fig.(\ref{higgstevlhc})   where the
 leading order
(LO) cross section for the sum of neutral Higgs processes
$\sigma_{\Phi \tau\tau} (p\bar p)=
[\sigma(p\bar p \to \Phi){\rm BR}(\Phi \to 2\tau)]$
(where sum over the neutral $\Phi$ fields is implied)
 vs the CP odd Higgs  mass is plotted for CM energy of  $\sqrt
s=1.96$ TeV  at the Tevatron. One finds  that the predictions of
$\sigma_{\Phi \tau\tau} (p\bar p)$ from the HPs are the largest and lie in a
narrow band  followed by those from the Chargino  Pattern mSP2.
The
recent data from the Tevatron is also shown\cite{Abazov:2006ih}. A
comparison of  the theory prediction with data shows that the HPs are
being constrained by experiment. Exhibited in  the right panel of
Fig.(\ref{higgstevlhc})  is
 $\sigma_{\Phi \tau\tau} (p p)=  [\sigma(p p \to \Phi){\rm BR}(\Phi \to 2\tau)]$
 arising from the HPs  (and also from other patterns which
make a comparable contribution) vs the CP odd Higgs  mass  with the
analysis done at  CM energy of  $\sqrt s=14$ TeV  at the LHC.  Again it is seen that
the predictions of  $\sigma_{\Phi \tau\tau} (p  p)$  arising
from the HPs are the largest and lie in a very narrow band
and the next largest predictions for
 $\sigma_{\Phi \tau\tau} (p  p)$ are typically from the Chargino Patterns (CPs).
The larger cross
sections for the HPs enhance the prospects of their detection.
Further, the analysis shows that the Higgs  production cross section
when combined with the parameter space  inputs and other signatures can be used to
discriminate  amongst mass patterns. Since the largest Higgs production cross sections
at the LHC
arise from the Higgs Patterns and the Chargino Patterns we exhibit the
mass of the light Higgs as a function of $m_0$ for these two patterns in
the left panel of Fig.(\ref{mandNUcross}). We note that many  of the Chargino Pattern
points in this figure appear to have large $m_0$ indicating that they originate from the
Hyperbolic Branch/Focus Point (HB/FP) region\cite{hb/fp}.

We discuss now briefly the Higgs to $b \bar b$ decay at the
Tevatron. From the parameter space of mSUGRA that enters in Fig.(1)
we can compute the quantity $[(p \bar p \to \Phi) {\rm
BR}(\Phi \to b \bar b)]$.
Experimentally, however, this quantity is
difficult to measure because there is a large background to the
production from $q\bar q, gg \to b \bar b$. For this reason one focuses on
the production $[(p \bar p \to \Phi b) {\rm BR}(\Phi \to b \bar
b)]$\cite{bh}.
For the parameter space of Fig.(1)   one gets
$[(p \bar p \to \Phi b)  {\rm
BR}(\Phi \to b \bar b)] \lesssim 1$~pb at ($\tan \beta = 55$,$M_A =
\rm 200~GeV$).
The preliminary CDF data \cite{CDFprelim} puts limits at
200~GeV, in the range (5-20)~pb over a $2 \sigma$ band at the tail
of the data set.
These limits are larger, and thus  less stringent,
than what one gets from  $\Phi\to \tau^+\tau^-$.
For the LHC, we find $[(p  p \to
\Phi b)  {\rm BR}(\Phi \to b \bar b)]\sim 200$~pb for the same
model point.  A more detailed fit requires a full treatment which is
outside the scope of the present analysis.

\begin{figure}[htbp]
  \begin{center}
\includegraphics[width=7.5cm,height=6.5cm]{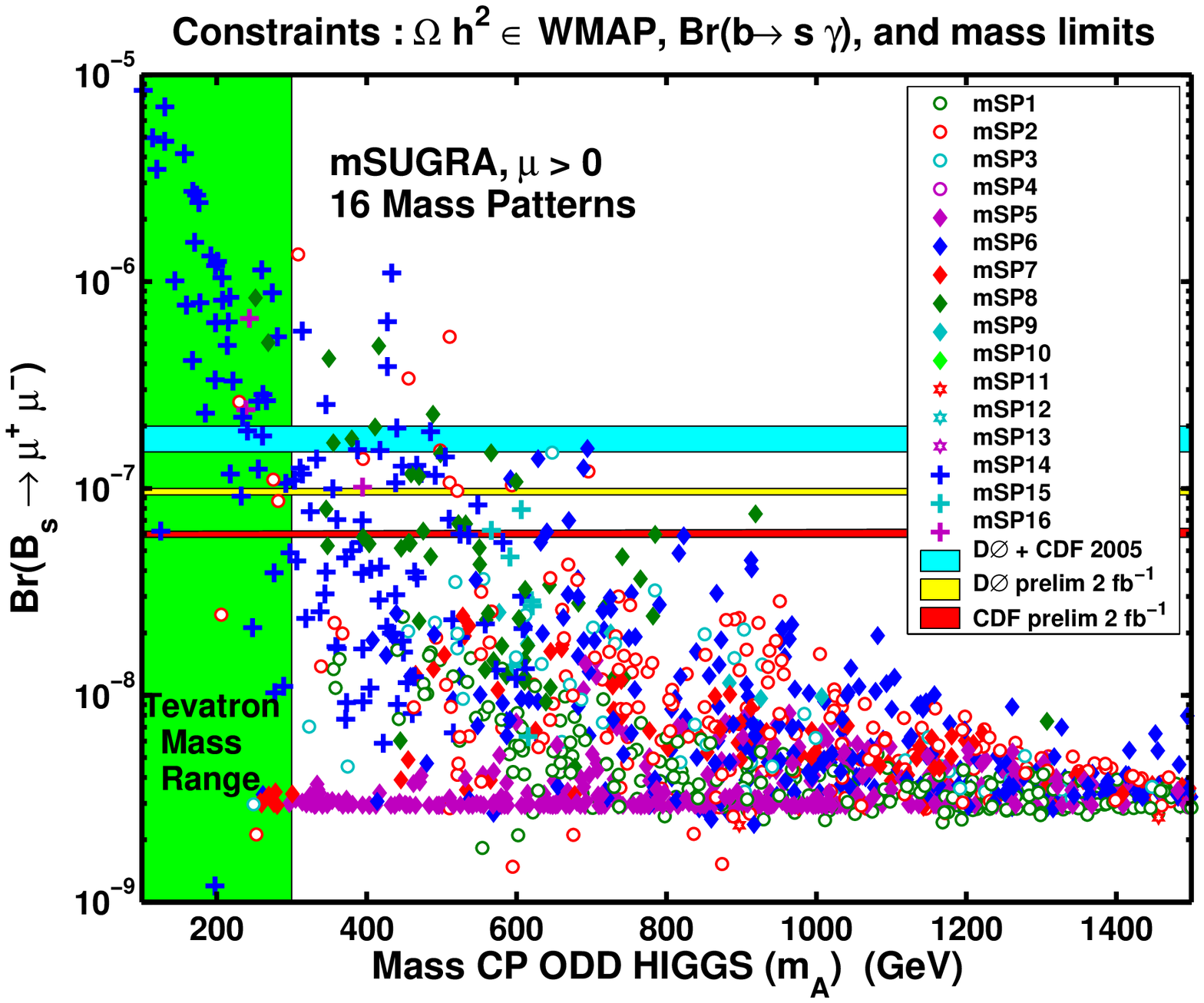}
\includegraphics[width=7.5cm,height=6.5cm]{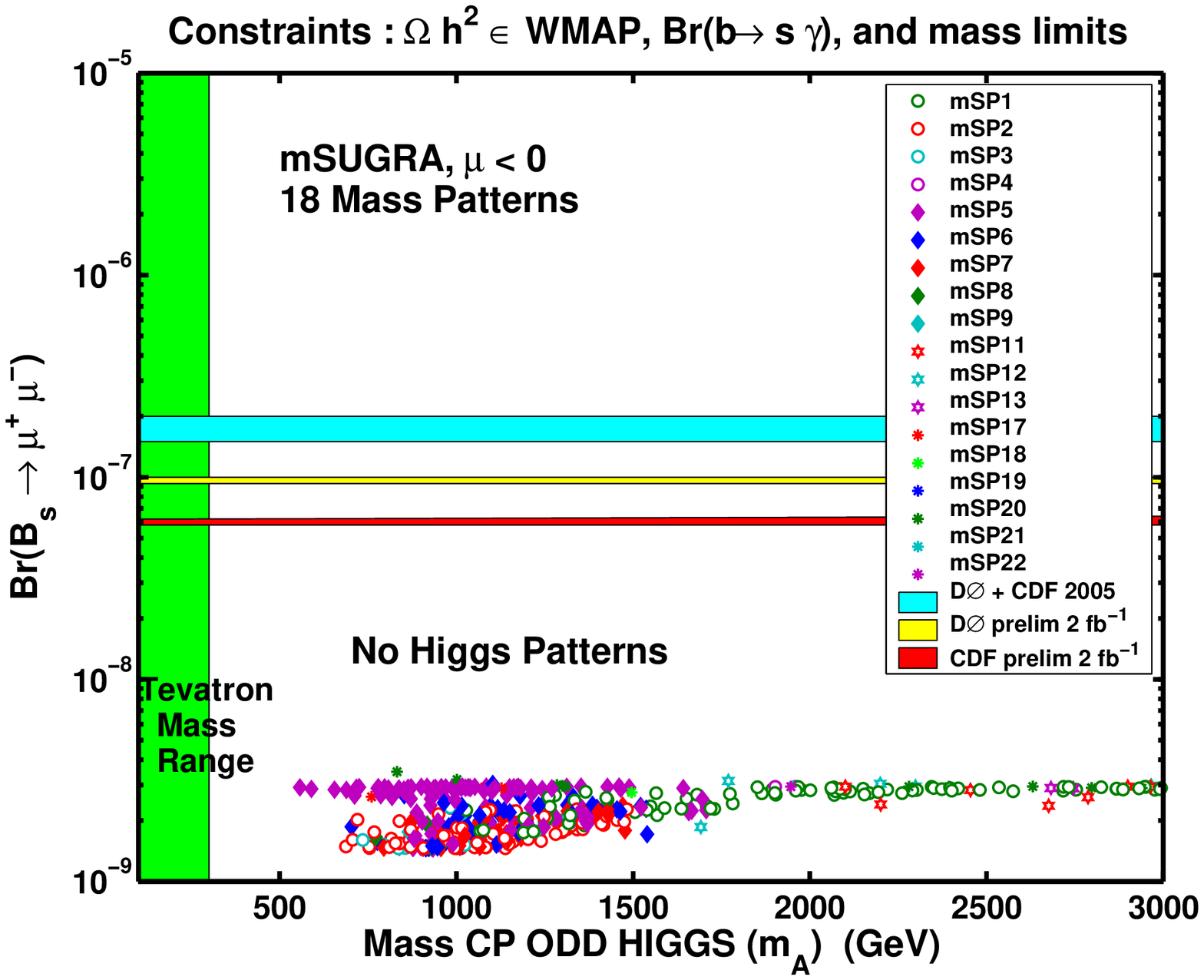}
\includegraphics[width=7.5cm,height=6.5cm]{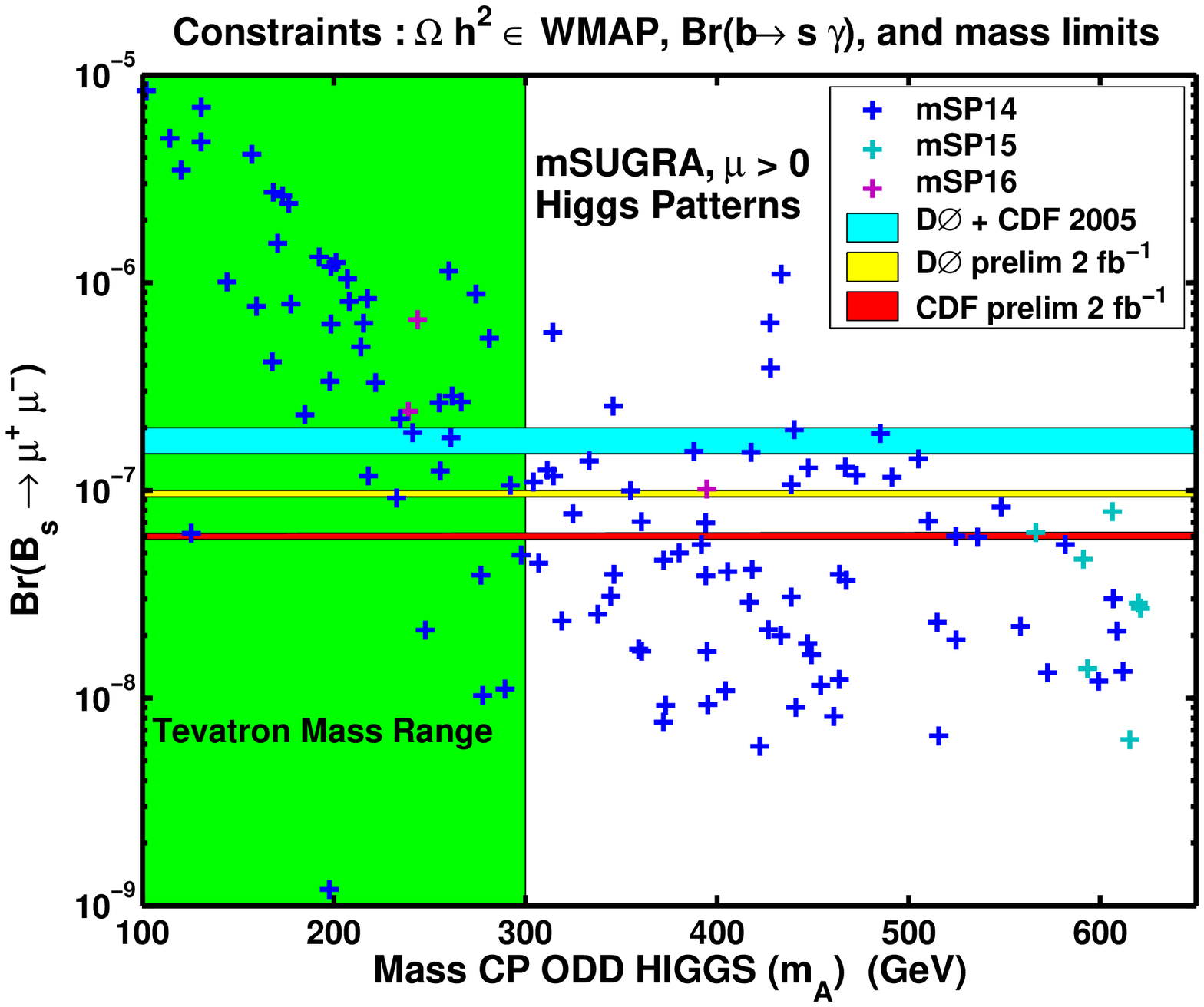}
\includegraphics[width=7.5cm,height=6.5cm]{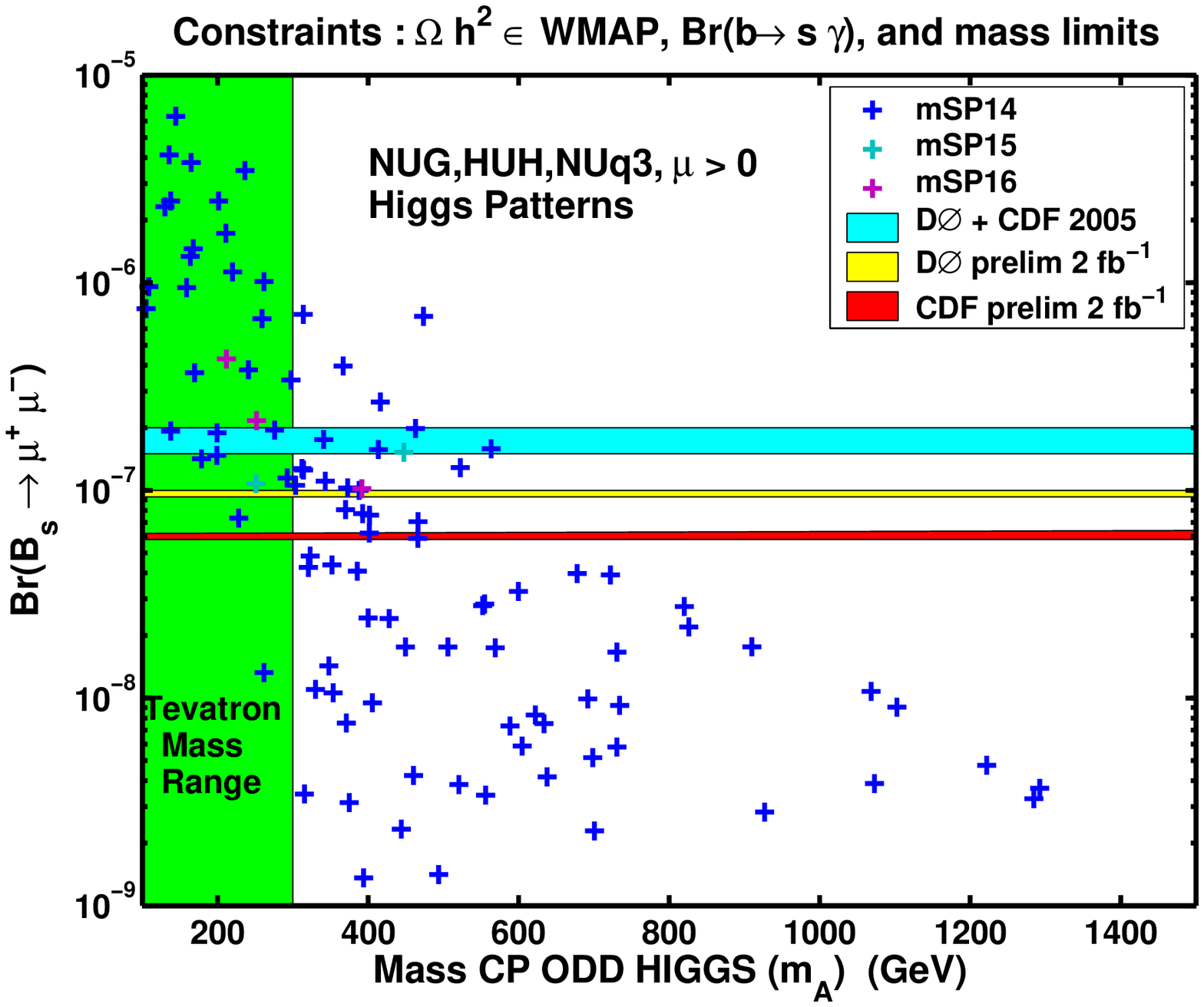}
\caption{(Color Online) Predictions for the branching ratio $B_s\to \mu^+\mu^-$ in various
patterns in the SUGRA landscape. Upper left  panel: predictions are for the patterns
for $\mu>0$ in mSUGRA; upper right panel: predictions are for the patterns for $\mu<0$ in mSUGRA;
lower left panel: predictions for the Higgs Patterns alone for $\mu>0$ in mSUGRA;
lower right panel: predictions
for NUSUGRA models NUH, NUq3, and NUG for $\mu>0$.
The experimental limits are: top band
2005 \cite{Bernhard:2005yn,Abulencia:2005pw}, and the bottom two horizontal lines
are preliminary limits from the CDF and D\O ~ data \cite{bsmumu07}.  For convenience  we draw the limits
extending past the  observable mass of the CP odd Higgs at the Tevatron.}
\label{bsmumu}
  \end{center}
\end{figure}

{\it $B_s\to \mu^+\mu^-$  and the Higgs  Patterns}: The process
$B_s\to \mu^+\mu^-$ is dominated by the neutral Higgs exchange
\cite{gaur}  and is enhanced by a factor of $\tan^6\beta$. It is
thus reasonable to expect that the HPs will be constrained more
severely than other patterns by the $B_s\to \mu^+\mu^-$
experiment,  since HP points usually arise from the high
$\tan\beta$ region (we note, however,  that the nonuniversalities in the  Higgs sector (NUH)
can also give rise to HPs for moderate  values of $\tan \beta$ (see Table(1))). This is supported
by a detailed analysis which is  given in
Fig.(\ref{bsmumu})  where the branching ratio ${\mathcal Br}(B_s\to
\mu^+\mu^-)$ is plotted against the CP odd Higgs mass $m_A$.  The
upper left (right)  hand panel gives the analysis for the case of mSUGRA for
$\mu>0$ ($\mu<0$)
 for the Higgs Patterns as well as for several other
patterns, and the experimental constraints are also shown.
One finds that the constraints are very effective for $\mu>0$ (but not for $\mu<0$)
constraining  a part of the parameter space of the HPs
and also some models within the Chargino and the Stau Patterns are
constrained (see upper left and lower left panels of Fig.(\ref{bsmumu})).

From the analysis of Fig.(3), it is observed that the strict imposition
of the constraint ${\mathcal Br } (B_s \to \mu^{+} \mu^{-}) < 1.5
\times 10^{-7}$ still allows for large  $\tan \beta$
in the mSUGRA model.  Thus all of the HP model points given
in Fig.(3) that satisfy this constraint  for the mSUGRA $\mu
> 0$ case correspond to $\tan \beta$ in the range of 50 - 55. A similar
limit on $\tan \beta$ is also observed for the nonuniversal models.
We remark, however, that the HPs are not restricted to large $\tan
\beta$ in particular for the case of the NUH model, where two such
benchmarks are given in Table(1) for quite moderate values of $\tan
\beta$. Here the HP model points in mSP14 for the NUH case in
Table(1) have ${\mathcal Br } (B_s \to \mu^{+} \mu^{-})\sim
(3.1,3.8) \times 10^{-9}$ which are significantly
lower than what is  predicted by the very large $\tan \beta$
case in models with universality and thus these cases are
much less constrained by the ${\mathcal Br } (B_s \to \mu^{+} \mu^{-}) $
limits.

\begin{figure}[htbp]
  \begin{center}
\includegraphics[width=7.5cm,height=6.5cm]{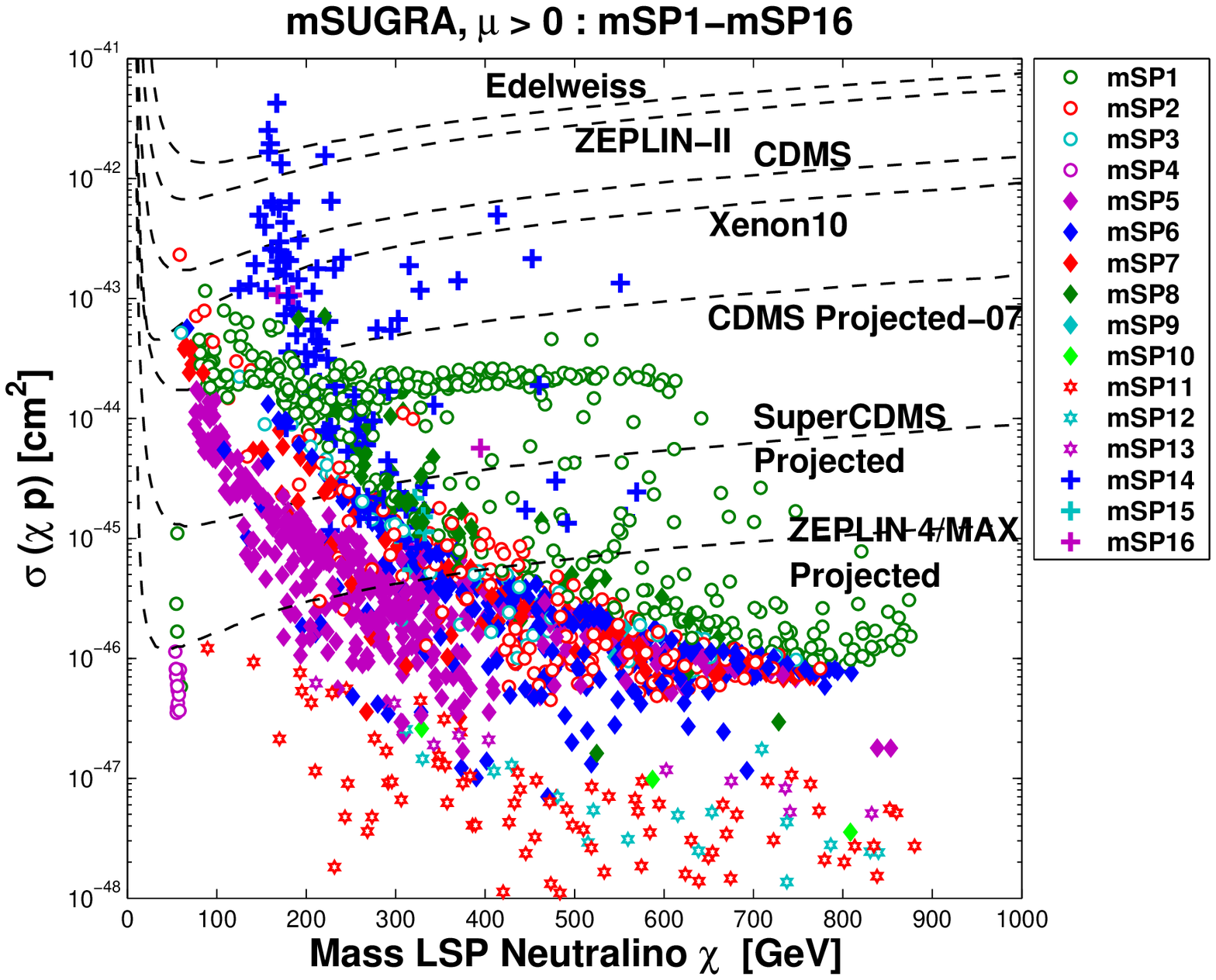}
\includegraphics[width=7.5cm,height=6.5cm]{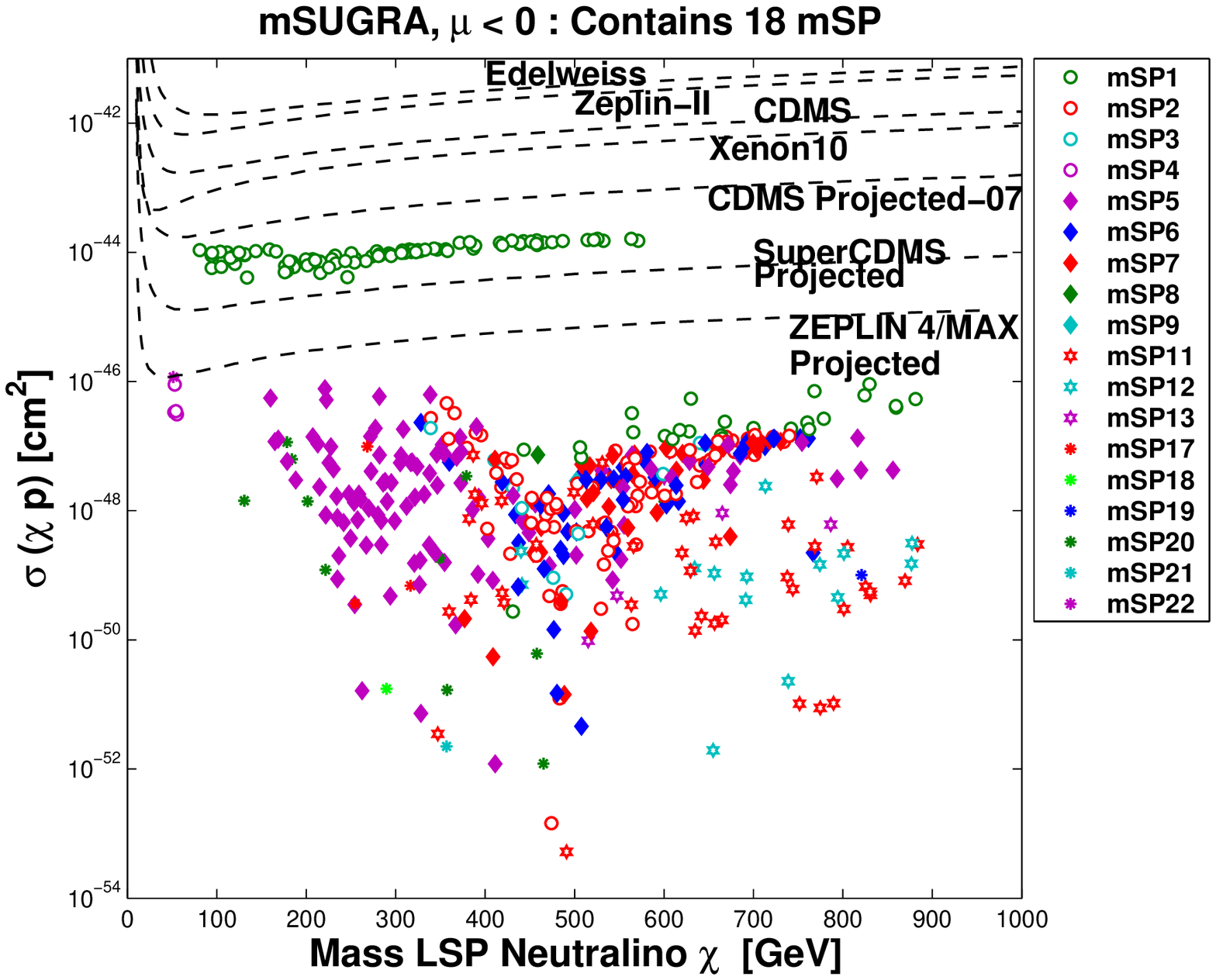}
\includegraphics[width=7.5cm,height=6.5cm]{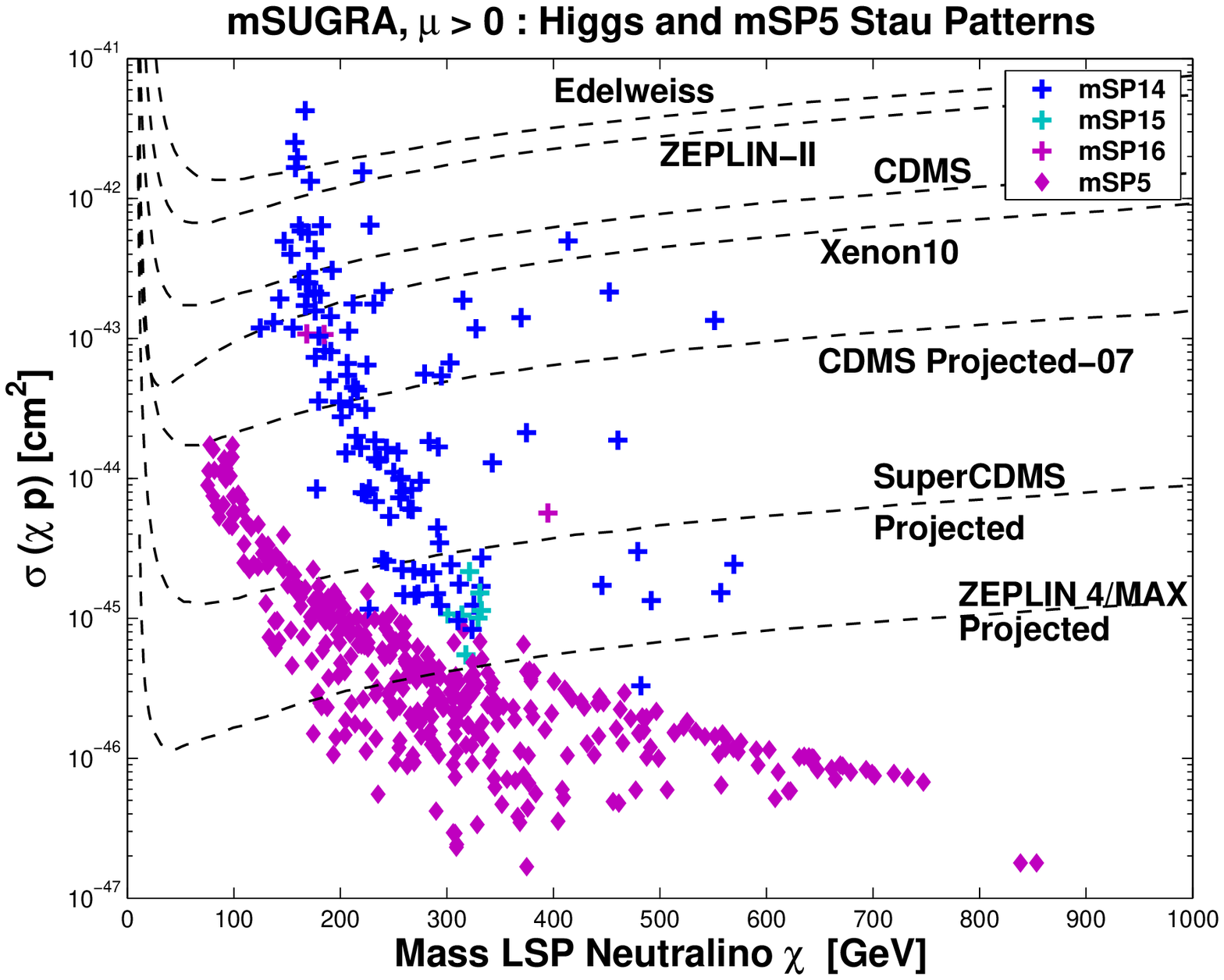}
\includegraphics[width=7.5cm,height=6.5cm]{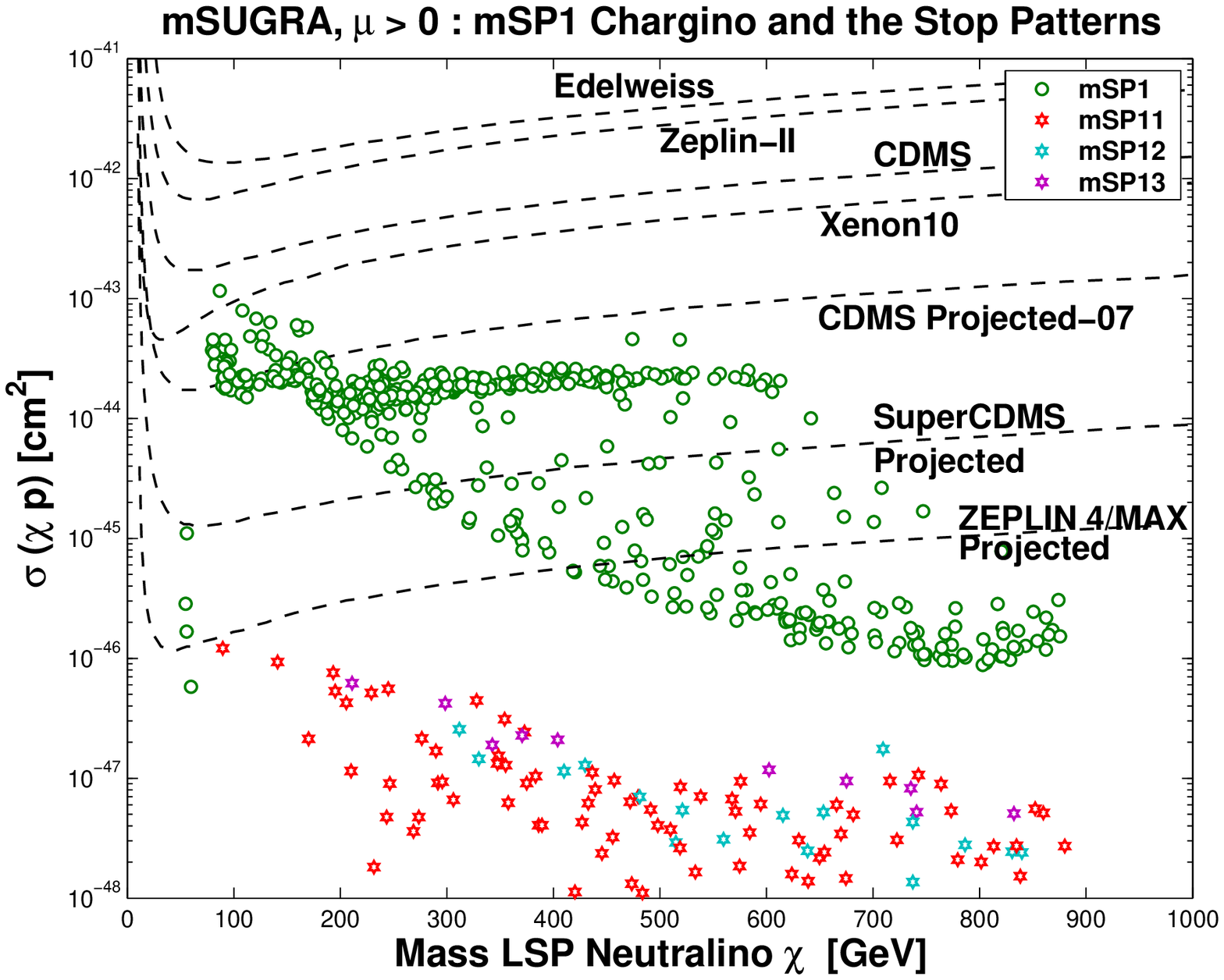}
\caption{(Color Online) Analysis of $\sigma(\chi p)$ for mSUGRA: upper left panel:
$\mu>0$ case including  all patterns; upper right panel:     $\mu<0$
allowing  all patterns; lower left hand  panel: A comparison of  $\sigma(\chi p)$ for HPs and a
stau NLSP case which is of type mSP5 for
$\mu>0$;
 lower right panel:  a comparison of  $\sigma(\chi p)$ for the Chargino Pattern
mSP1 vs  the Stop Patterns mSP11-mSP13. The analysis shows  a
Wall  consisting of a clustering of points in the Chargino
Patterns mSP1-mSP4
with a $\sigma(\chi p)$ in the range $10^{-44\pm .5}$ cm$^2$
enhancing the prospects for the observation of dark matter by
SuperCDMS \cite{Schnee:2005pj},  ZEPLIN-MAX\cite{Atac:2005wv}
or LUX\cite{lux} in this region.}
\label{dcross}
  \end{center}
\end{figure}

 {\it Dark Matter-Direct Detection}: We discuss now the
direct detection of dark matter. In Fig.(\ref{dcross}) we give an
analysis of the scalar neutralino-proton cross  section
$\sigma({\tilde\chi_1^0 p})$ as a function of the LSP mass (complete analytic
formulae for the computation of dark matter cross sections can be
found in \cite{Chattopadhyay:1998wb} and for a sample of  Post-WMAP3
analysis of dark matter see \cite{modern,modern1}). The upper  left
panel of Fig.(\ref{dcross}) gives the scalar $\sigma({\tilde\chi_1^0 p})$ for
the
 mSUGRA parameter space for $\mu>0$.
We  note  that the Higgs  patterns  typically give the largest dark matter cross
sections (see  the upper left and lower left panels of Fig.(\ref{dcross}))
and are the first ones to be constrained by experiment.
The second largest cross sections arise from the Chargino Patterns
which shows an embankment, or Wall, with a copious number of points with
cross sections in the range $10^{-44\pm .5}$cm$^2$ (see the upper left panel and
lower right panel), followed by Stau Patterns (lower left panel), with the
Stop Patterns producing the smallest cross  sections (upper left and lower right panels).
The upper right panel of Fig.(\ref{dcross}) gives the scalar  cross section $\sigma({\tilde\chi_1^0 p})$
for $\mu<0$ and here one finds  that the largest cross  sections arise  from the CPs
which also have a Chargino Wall with cross sections in the range $10^{-44\pm .5}$cm$^2$
(upper right panel).
The  analysis shows that altogether
the scalar  cross sections lie in an
interesting region and would be accessible to dark matter  experiments
currently underway and improved experiments  in the
future \cite{Bernabei:1996vj,Sanglard:2005we,Akerib:2005kh,Alner:2007ja,Angle:2007uj,lux}.
Indeed  the analysis of Fig.(\ref{dcross}) shows that some of the parameter
space of the Higgs Patterns is beginning to be constrained by the CDMS
and the Xenon10 data \cite{Angle:2007uj}.

What is very  interesting is the fact that  for the case $\mu>0$  the  $B_s\to \mu^+\mu^-$ limits, the
Tevatron limits on  the CP odd Higgs boson production,  and the CDMS and Xenon10  limits
  converge on constraining the Higgs Patterns and specifically the pattern  mSP14 and as well as
some other patterns.
 Thus the  CDMS and Xenon10  constraints on the mSPs are  strikingly similar to the constraints
of $B_s\to\mu^+\mu^-$ from the Tevatron.
 We also observe that although  the case $\mu<0$ is not currently accessible to the
 $B_s\to \mu^+\mu^-$ constraint (and may  also be beyond the ATLAS/CMS sensitivity for $B_s\to\mu^+\mu^-$),
 it would, however, still be accessible at least partially to dark matter
 experiment.  Finally we remark that the proton-neutralino cross
 sections act as a discriminator of the SUGRA patterns as it  creates a significant
 dispersion among some of the patterns
 (see upper  left and  the two  lower  panels  in Fig.(\ref{dcross})).

{\it Nonuniversalities  of soft breaking}: Since the nature of
physics at the Planck scale is largely unknown it is useful to
consider other soft breaking scenarios beyond mSUGRA. One such
possibility is to consider nonuniveralities in the K$\ddot{\rm a}$hler potential,
which can give rise to nonuniversal soft breaking consistent with
flavor changing neutral current constraints.  We consider three
possibilities  which are nonuniversalities in    (i)   the Higgs
sector (NUH), (ii) the third  generation squark sector (NUq3), and
(iii)  the gaugino sector (NUG) (for a sample of previous work on
dark matter analyses with nonuniversalities see \cite{Nath:1997qm}).
We parametrize these  at the GUT scale as follows: (i) NUH: $M_{H_u}
= m_0(1+\delta_{H_u})$,  $M_{H_d} = m_0(1+\delta_{H_d})$, (ii) NUq3:
$M_{q3}=m_0(1+\delta_{q3})$,  $M_{u3,d3}=m_0(1+\delta_{tbR})$, and,
(iii) NUG: $M_{1}=m_{1/2}$, $M_{2}=m_{1/2}(1+\delta_{M_2})$,
$M_{3}=m_{1/2}(1+\delta_{M_3})$, with $-0.9\leqslant\delta\leqslant
1$. In each case we carry out a Monte Carlo scan of  $1 \times 10^6$  models. The above
covers  a  very   wide array of models.
 The analysis here  shows that the patterns that appear in
 mSUGRA  (i.e., mSPs) also appear here.  However, in addition to the mSPs, new
 patterns  appear which are labeled NUSP1-NUSP15  (see Table(\ref{tab:nusugra})),  and
 we note the appearance of gluino patterns, and patterns where both the Higgses and gluinos
 are among the lightest sparticles.
 The neutral Higgs production cross section for the NUSUGRA case is given in the right panel of
  Fig.(\ref{mandNUcross}).  The analysis shows that the Higgs Patterns produce  the largest
  cross sections followed by the Chargino  Patterns as in mSUGRA case.
 The constraints of ${\mathcal Br}(B_s\to \mu^+\mu^-)$ on the NUSUGRA Higgs patterns are
 exhibited  in the lower right hand panel of   Fig.(\ref{bsmumu}).
Again one  finds that the ${\mathcal Br}(B_s\to \mu^+\mu^-)$ data constrains the parameter
space of the HPs in the NUSUGRA case.
 One feature which is now different is that the Higgs Patterns survive
significantly beyond the CP odd Higgs  mass of 600 GeV within our assumed naturalness
assumptions. Thus
nonuniversalities tend to extend the CP odd Higgs beyond what one
has in the mSUGRA case.

\begin{table}[htbp] \begin{center}
 \scriptsize{
\begin{tabular}{||c||c||c|c||}
\hline\hline NUSP Label &   Pattern &     NUSP Label  &  Pattern
\\\hline\hline
 {\bf NUSP1}   &   $\na$   $<$ $\cha$  $<$ $\nb$   $<$ $\ta$  & {\bf NUSP2}  &   $\na$ $<$ $\cha$  $<$ $A,H$\cr
 {\bf NUSP3}     &   $\na$   $<$ $\cha$ $<$ $\sta$ $<$ $\nb$    & {\bf NUSP4}  & $\na$ $<$ $\cha$ $<$ $\sta$  $<$ $\slr$\cr
 {\bf NUSP5}    &   $\na$   $<$ $\sta$  $<$ $\snl$  $<$ $\stb$ & {\bf NUSP6}  & $\na$   $<$ $\sta$ $<$ $\snl$  $<$ $\cha$\cr
 {\bf NUSP7}    &   $\na$ $<$ $\sta$  $<$ $\ta$   $<$ $A,H$    & {\bf NUSP8} & $\na$   $<$ $\sta$ $<$ $\slr$ $<$ $\snm$ \cr
 {\bf NUSP9}     &   $\na$ $<$ $\sta$ $<$ $\cha$  $<$ $\slr$    & {\bf NUSP10} &   $\na$ $<$ $\ta$ $<$ $\g$ $<$ $\cha$ \cr
 {\bf NUSP11}    &   $\na$ $<$ $\ta$   $<$ $A,H$                & {\bf NUSP12} & $\na$ $<$ $A,H$ $<$ $\g$  \cr
 {\bf NUSP13}    & $\na$ $<$ $\g$    $<$ $\cha$ $<$ $\nb$       & {\bf NUSP14}&  $\na$   $<$ $\g$    $<$ $\ta$ $<$ $\cha$ \cr
{\bf NUSP15}    &   $\na$   $<$ $\g$ $<$ $A,H$ & &  \cr
  \hline\hline
\end{tabular}
} \caption{ New 4 sparticle mass patterns for NUSUGRA in a $3 \times
10^6$ model scan of the parameter space with nonuniversalites. The
new patterns are  labeled NUSP1-NUSP15. NUSP(1,2,5,6) appear in NUq3 and NUSP(1,3,4;7-15) appear in NUG.
NUH contains only mSPs.}
 \label{tab:nusugra}
\end{center}
 \end{table}

\begin{figure}[htbp]
  \begin{center}
\includegraphics[width=7.5cm,height=6.5cm]{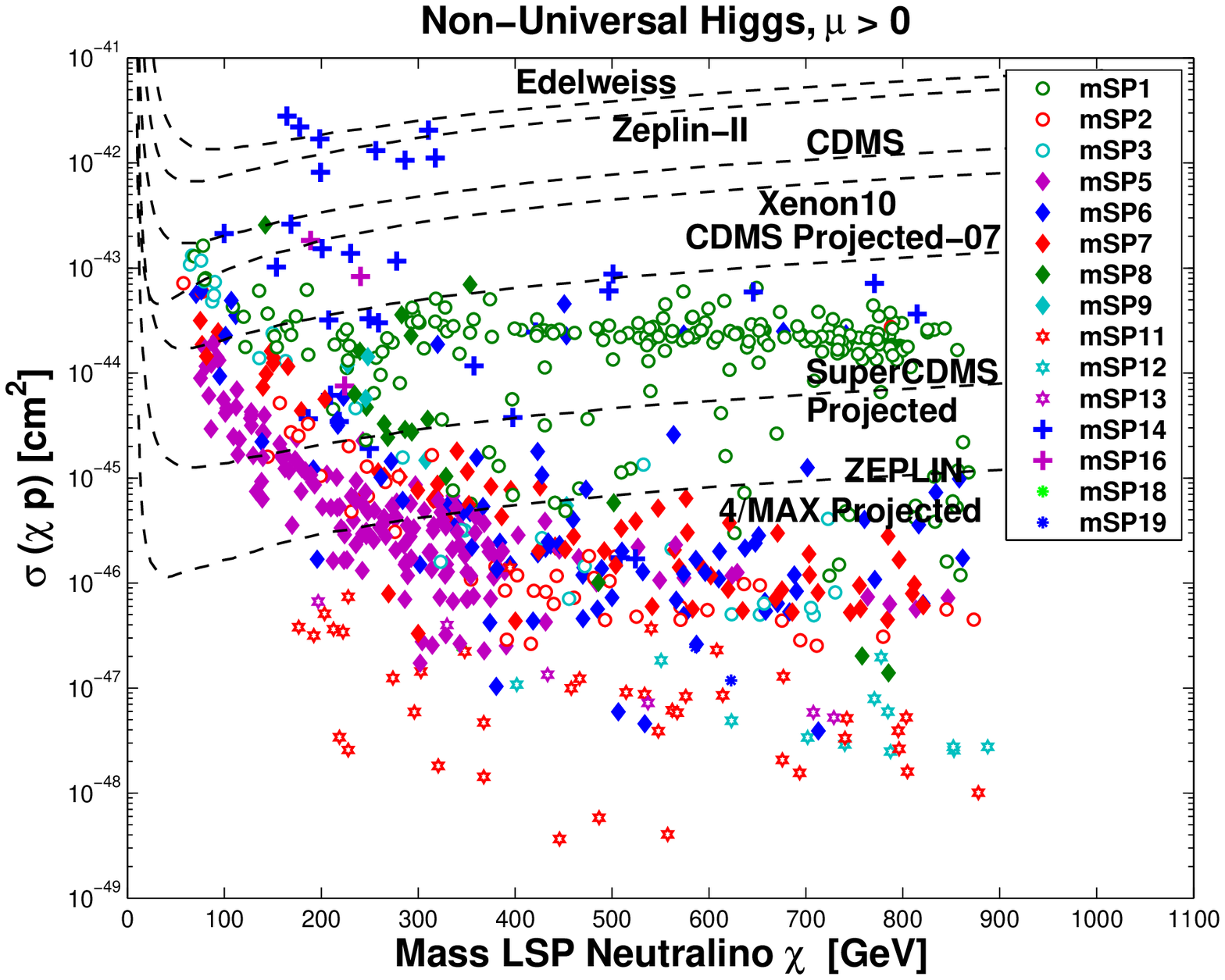}
\includegraphics[width=7.5cm,height=6.5cm]{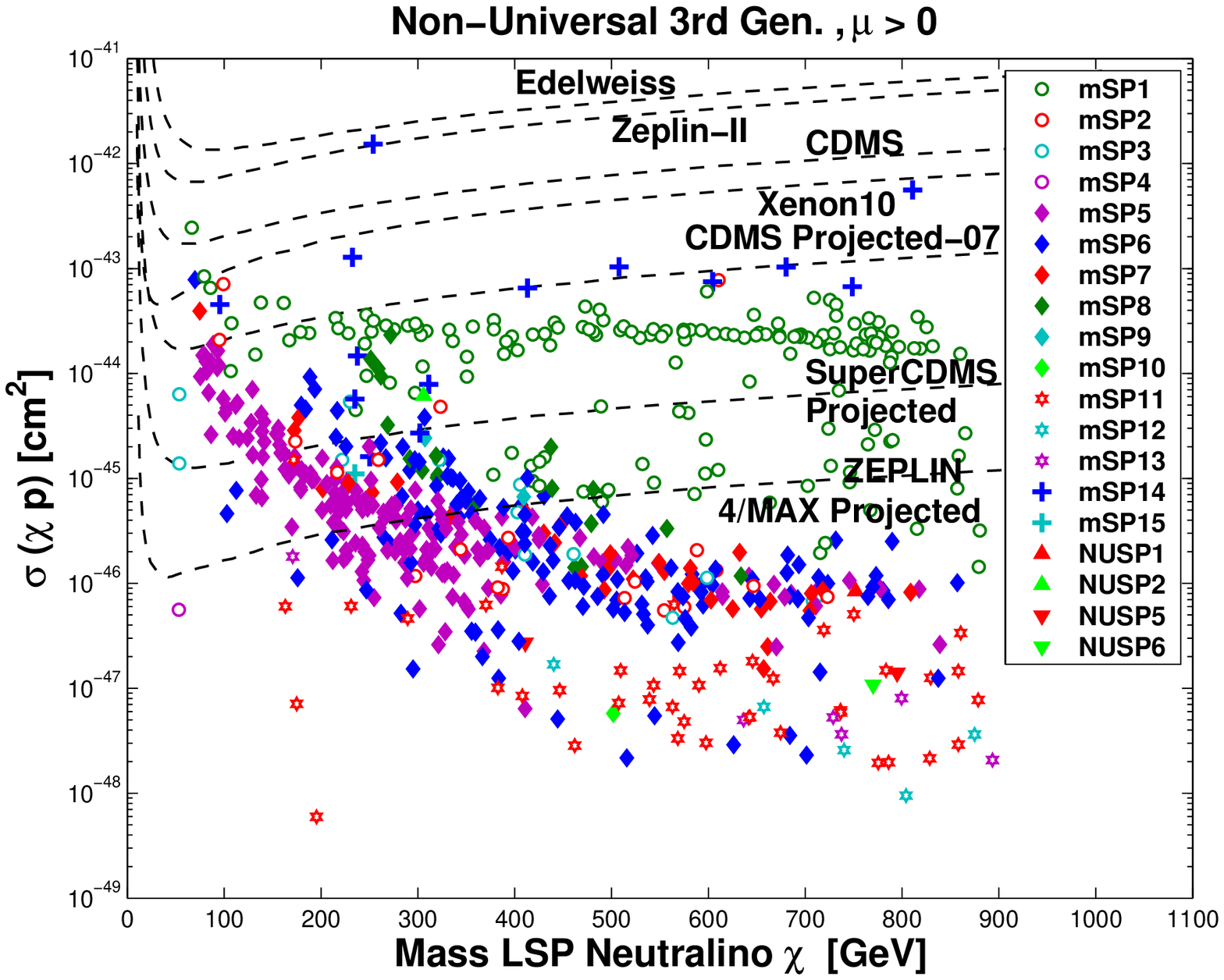}
\includegraphics[width=7.5cm,height=6.5cm]{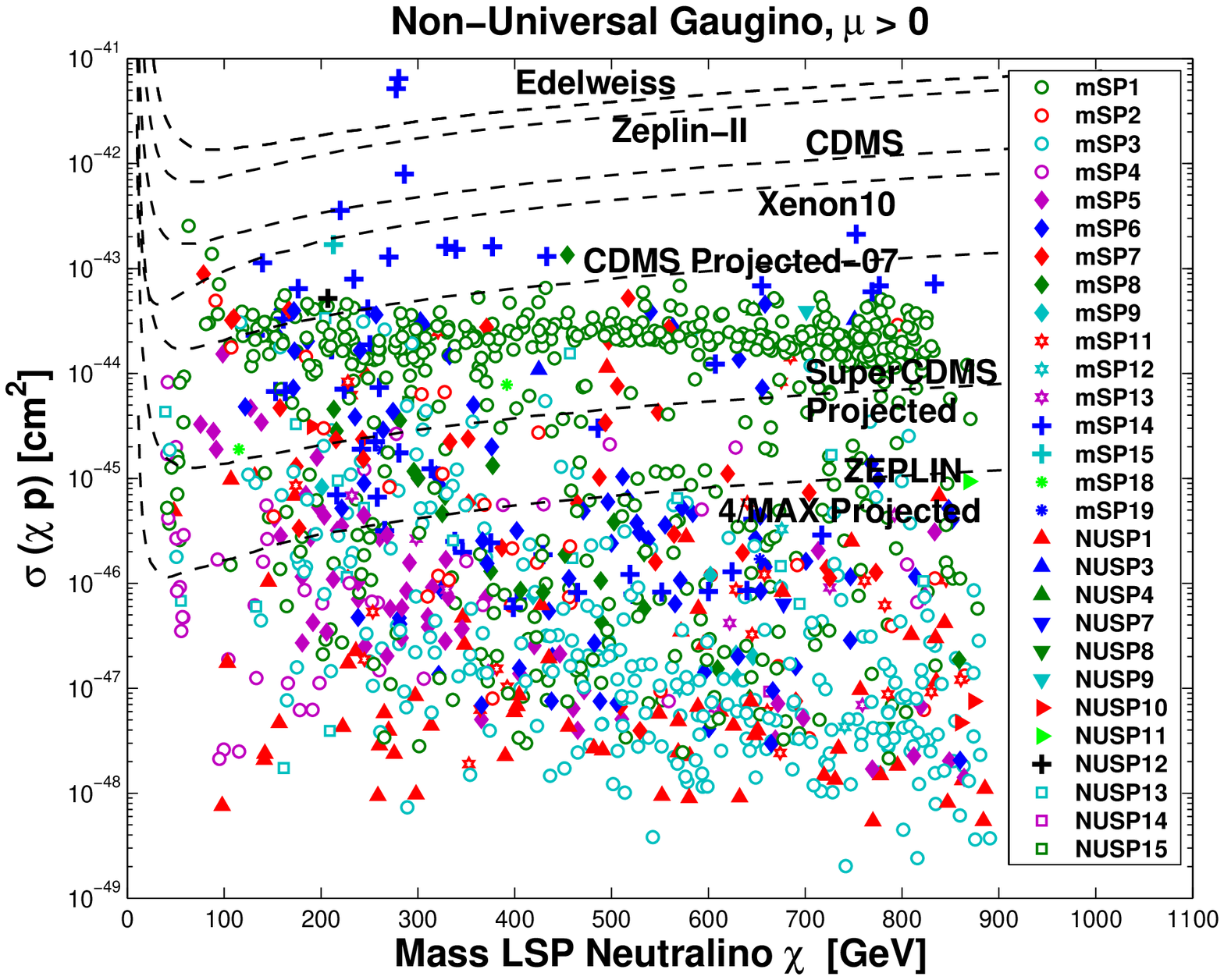}
\includegraphics[width=7.5cm,height=6.5cm]{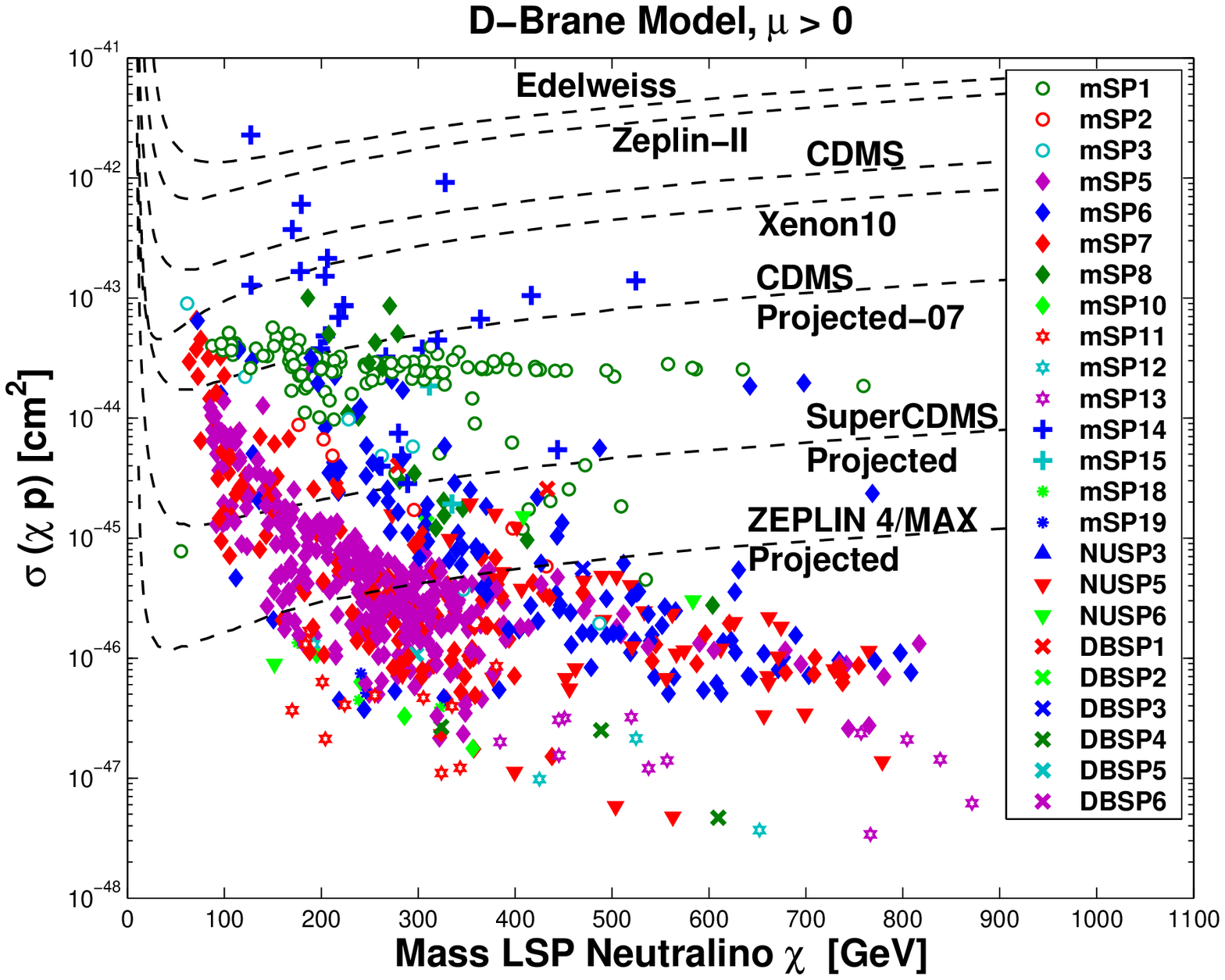}
\caption{(Color Online) Analysis of the scalar cross section $\sigma(\chi p)$ for
NUSUGRA  and D-brane models:  NUH (upper left panel), NUq3 (upper right  panel),
NUG  (lower left panel), and brane models (lower right panel).  As in Fig.(\ref{dcross})
the  Wall consisting
of a clustering of points in the Chargino Patterns mSP1-mSP4  persists  up to an LSP mass of
about 900 GeV with a $\sigma(\chi p)$ in the range $10^{-44\pm .5}$ cm$^2$
enhancing the prospects for the observation of dark matter by SuperCDMS and ZEPLIN-MAX
in this region.}
\label{dnonuni}
  \end{center}
\end{figure}

  Next we analyze  the direct detection of  dark matter in NUSUGRA.
The results of the analysis are presented in
Fig.(\ref{dnonuni}) (upper two panels  and the lower left panel).
As in the mSUGRA case one finds
that the largest dark matter cross sections still arise from
the Higgs Patterns followed  by the Chargino Patterns
within the three types of nonuniversality models considered:
 NUH (upper left panel   of Fig.(\ref{dnonuni})),
 NUq3 (upper right panel   of  Fig.(\ref{dnonuni})), NUG (lower left panel
of Fig.(\ref{dnonuni})).
Again
the analysis within  NUSUGRA
shows the phenomenon of the Chargino Wall, i.e., the existence of a
copious  number of
Chargino  Patterns (specifically mSP1) in all cases with cross sections in the
range $10^{-44\pm .5}$cm$^2$.
Most of the parameter  points  along the Chargino
Wall lie on the Hyperbolic Branch/Focus Point (HB/FP) region\cite{hb/fp}
where the Higgsino components of the LSP are substantial (for a review see \cite{Lahanas:2003bh}). Thus this
Chargino Wall presents an encouraging region of the parameter space
where the dark matter may become observable in improved experiments.

{\it Light Higgses and Dark Matter in D-brane Models:}  The advent
of D-branes has led to a new wave of model building
\cite{dreviews},  and several Standard Model like extensions have
been constructed using intersecting  D-branes \cite{dmodels}. The
effective action and soft breaking in such models have been discussed
\cite{dsoft} and there is some progress also on pursuing the
phenomenology of intersecting D-brane models
\cite{dpheno,pheno1,Cvetic:2002qa}.  Here we discuss briefly the
Higgses and dark matter in the context of D-branes.
In our analysis we use the scenario of
toroidal  orbifold compactification based on  ${\cal
T}^6/\mathbb{Z}_2\times \mathbb{Z}_2$ where ${\cal T}^6$ is taken
to be  a product of 3  ${\cal T}^2$ tori.   This model has a moduli
sector consisting of volume moduli $t_m$,  shape moduli $u_m$  $(m
=1,2,3)$  and the axion-dilaton field $s$. The detailed form of the
soft breaking in D-brane models
can be found in \cite{dsoft}, and we focus here on the
$\frac{1}{2}$ BPS sector.
Specifically the parameter space consists
of the gravitino mass $m_{3/2}$, the gaugino mass $m_{1/2}$, the
trilinear coupling $A_0$,  $\tan\beta$, the stack angle $\alpha$
($0\leqslant \alpha \leqslant \frac{1}{2}$), the  Goldstino
angle \cite{Brignole:1997dp} $\theta$,
the moduli  VEVs, $\Theta_{t_i}$, $\Theta_{u_i}$  $(i=1,2,3)$
obeying the sum rule $\sum_{i=1}^3 F_i=1$, where
 $F_i=  |\Theta_{t_i}|^2+|\Theta_{u_i}|^2$, and sign($\mu$)
(see
Appendix  A of  the first paper in \cite{dsoft} for details).  In
the analysis we ignore the exotics, set $F_3=0, 0\leqslant F_1\leqslant 1$,
   and use
 the naturalness assumptions similar
to the mSUGRA case with $\mu>0$.   The analysis shows that the allowed  parameter
space  is dominated by
the mSPs with only six new patterns (at isolated points ) emerging. Specifically  all the HPs (mSP14-mSP16) are
seen to emerge in good abundance.  Regarding the new  patterns
we label these patterns D-brane Sugra Patterns (DBSPs) since the
patterns arise in the SUGRA field point limit of the D-branes. Specifically we find
six new patterns  $\rm DBSP(1-6)$ as follows \begin{equation}
\begin{array}{ll}
{\rm DBSP1:}  ~\na < \sta < \snl  <A/H ~;
&{\rm DBSP2:}  ~\na < \sta < \snl <\slr ~;\\
{\rm DBSP3:}  ~\na < \sta < \snl  <\snm ~;
&{\rm DBSP4:}  ~\na < \ta  < \sta  <\snl ~;\\
{\rm DBSP5:}  ~\na < \snl  < \sta <\snm ~;
&{\rm DBSP6:}  ~\na < \snl  < \sta <\cha ~.
\end{array}
\end{equation}
\begin{table}[h]
    \begin{center}
    \scriptsize{
\begin{tabular}{|c|c|c|c|c|c|c|c|}
                                                                    \hline  \hline \hline
              &   $m_{3/2}$   &   $m_{1/2}$   &   $A_{0}$ &   $\tan {\beta}$  &     $\alpha$ &   $\cos^2\theta$ &   $F_1$ \\
{\bf\rm DBSPs}  &   (GeV)   &   (GeV)   &   (GeV)   &   $  $   &      &    &                  \\ \hline  \hline
 ${\bf DBSP1 }$  &  3654  &  1018     &    -331   &   51.5    &   0.444     &  0.705   &   0.086    \\ \hline
 ${\bf DBSP4 }$  &  1962  &  777      &    5863   &   9.4     &   0.430     &  0.790   &   0.260    \\ \hline
 ${\bf DBSP5 }$  &  2114  &  718      &    3512   &   21.3    &   0.448     &  0.688   &   0.051    \\ \hline
                                                                    \hline \hline
                                                                        \end{tabular}
}
\caption[]{Benchmarks for D-brane models
DBSPs.}
 \label{table:dbsp}
    \end{center}
 \end{table}

\begin{figure}[htbp]
  \begin{center}
\includegraphics[width=7.5cm,height=6.5cm]{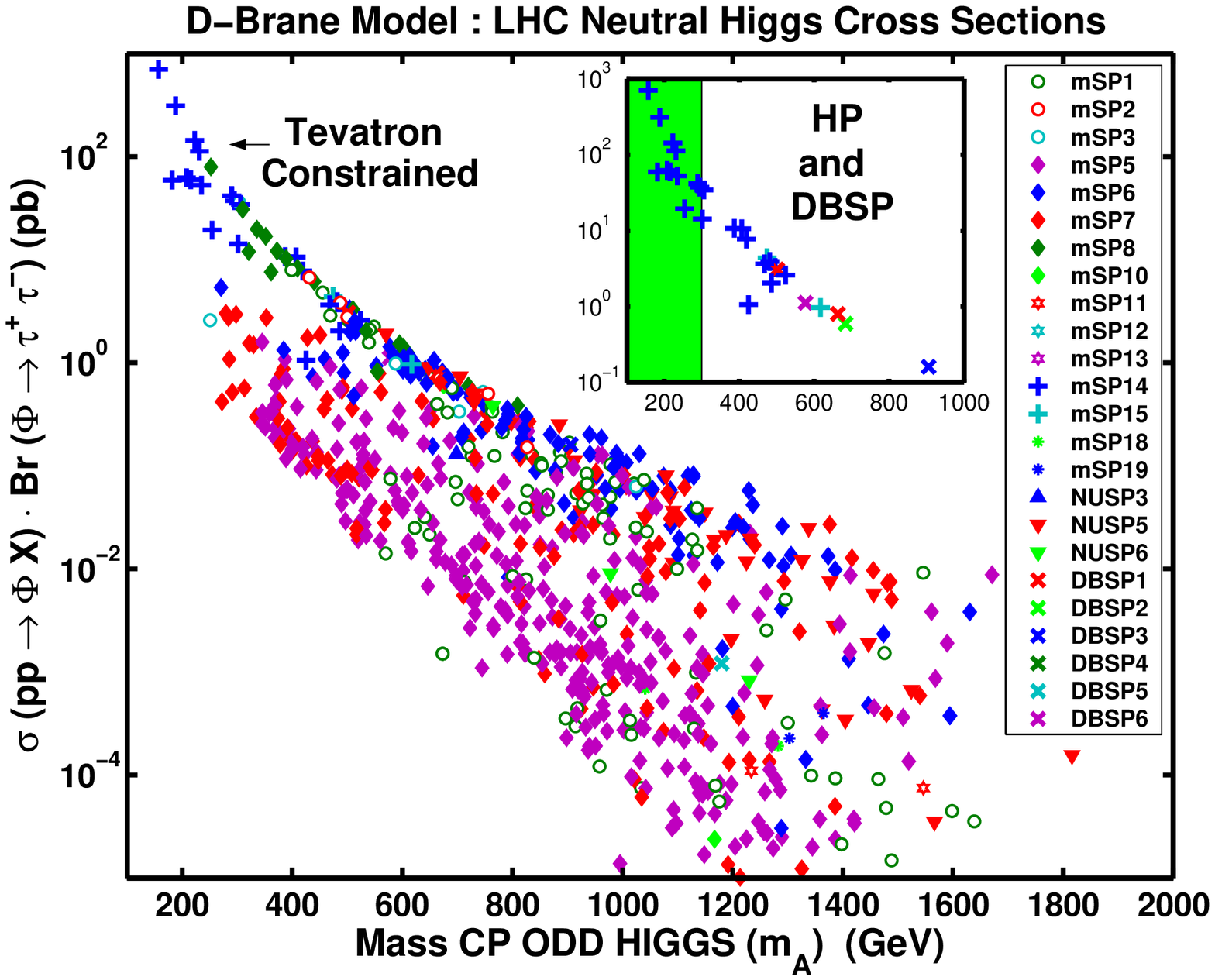}
\includegraphics[width=7.5cm,height=6.5cm]{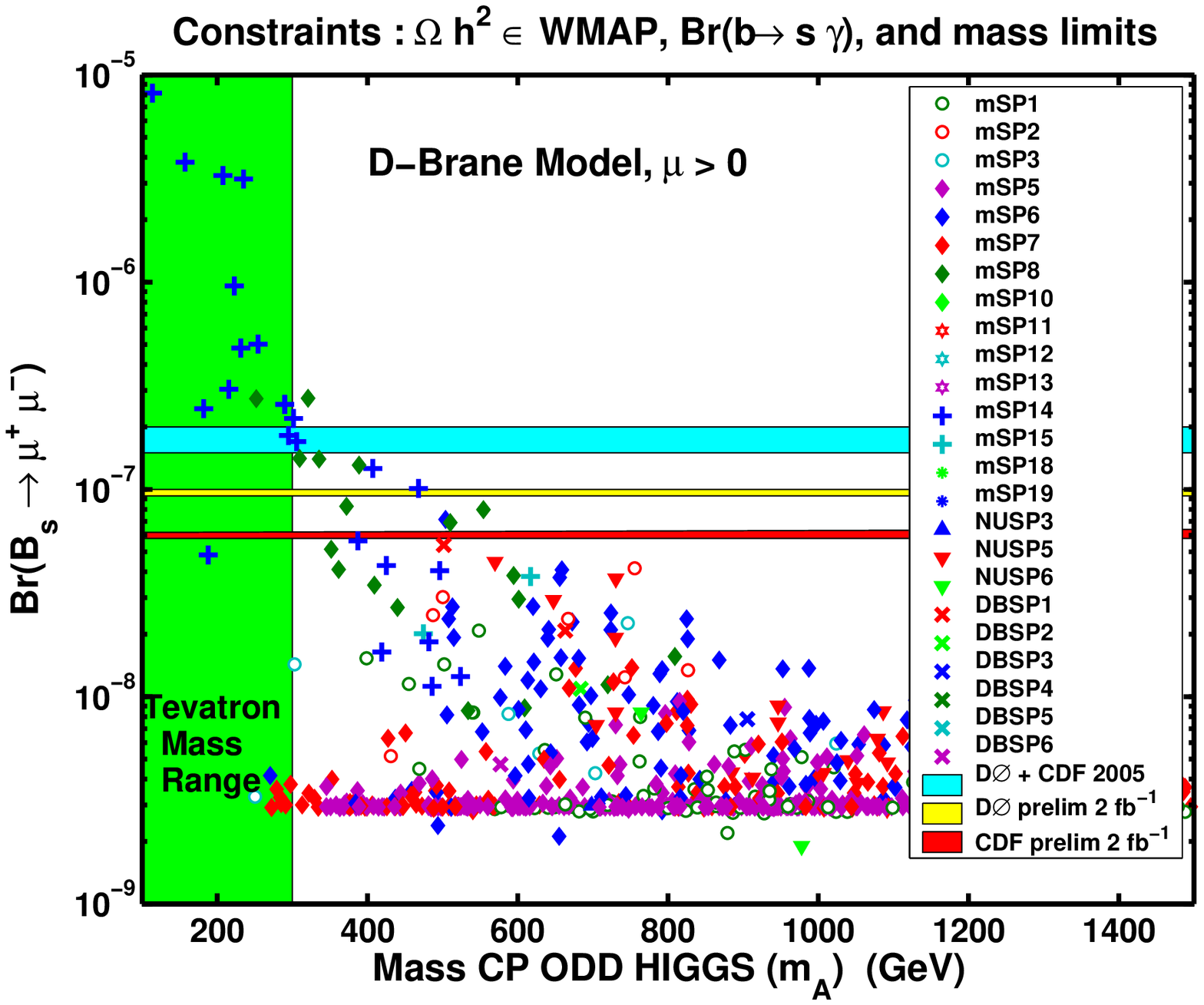}
\caption{(Color Online) Predictions  in D-brane models for $\mu>0$:
The Higgs  production cross section $\sigma_{\Phi\tau\tau}(pp)$ at the LHC as a
function of the CP odd  Higgs mass $m_A$ (left panel);
  $B_s\to \mu^+\mu^-$ vs  $m_A$ (right panel). The experimental constraints  from the
  Tevatron are  shown and constrain the Higgs  Patterns.}
\label{KN}
  \end{center}
\end{figure}

\noindent The analysis of the Higgs  production cross section
$\sigma_{\Phi \tau \tau}(pp)$ in the D-brane models  at the LHC is
given in the left panel of Fig.(\ref{KN}).
 The
analysis shows that the HPs again dominate the Higgs production
cross sections.  One also finds that the $B_s\to \mu^+\mu^-$ experiment constraints
the HPs in this model as  seen in the right panel of Fig.(\ref{KN}).
The dark matter  scalar  cross  section $\sigma(\na p)$ is given in the lower  right panel of
Fig.(\ref{dnonuni}).  Here also one finds  that the Higgs Patterns  typically give the largest
scalar cross sections followed by the Chargino Patterns (mSP1-mSP3)
 and then by  the Stau Patterns.
Further, one finds that the Wall  of  Chargino
Patterns persists in this case as well.

We comment briefly on the signals from the chargino patterns. The
chargino patterns correspond typically to low values of $m_{1/2}$
and arise dominantly from the hyperbolic branch/focus point region
of radiative breaking of the electroweak symmetry.
The above situation then gives rise to light
charginos and neutralinos which can produce a copious number of
leptonic signatures. We note that in the recent analysis of Ref.\cite{Feldman:2007zn},
the chargino pattern was studied in detail and the signatures at the LHC
investigated.
In particular  it is found that the chargino patterns can give
rise to substantial di-lepton and tri-leptonic signatures. Thus suppose
 we consider a model point in mSUGRA $\mu >0$ that sits on
the  Chargino Wall with $(m_0,
m_{1/2},A_0,\tan\beta)=(885,430,662,50.2)$ (mass units in GeV). Here
$(m_{\na},m_{\cha})\sim (177,324)~\rm GeV$ with $\sigma(\na p)\sim
1\times 10^{-8}~\rm pb$, and $\Omega_{\na}h^2 \sim .085$.
An analysis of leptonic signatures at the LHC  with $10~\rm fb^{-1}$
in this case gives  the number of  di-lepton
and tri-lepton SUSY events ($N$)  with the cuts imposed as in Ref.\cite{Feldman:2007zn},
so that $(N_{2L},N_{3L})_{jet
\geq 2}\sim (350,40)$, (where ($L=e,\mu)$). Both signatures are
  significantly  above the $5\sigma$ discovery limits at the LHC (see Ref.\cite{Feldman:2007zn}).

{\it Conclusions}: It is seen that Higgs Patterns (HPs) arise in a
wide range of models: in mSUGRA, in NUSUGRA and in D-brane models.
The HPs are typically  seen to lead to large Higgs production cross
sections at the Tevatron and at the LHC, and  to the largest $B_s\to
\mu^+\mu^-$ branching ratios, and thus are the first to be
constrained by the  $B_s\to \mu^+\mu^-$ experiment. It is also seen
that the HPs lead typically to the largest neutralino-proton cross
sections and would either be the first to be observed or the first
to be constrained by dark matter experiment.
The analysis presented here shows the existence of
 a Chargino Wall consisting of a copious number of parameter  points
 in the Chargino Patterns where
the NLSP is a chargino which give a $\sigma(\tilde\chi_1^0 p)$  at the level of $10^{-44\pm .5}$cm$^2$
in all models considered for the LSP mass extending up to 900 GeV in many cases.
These results  heighten the possibility for the observation of dark matter in improved dark
matter experiments such as  SuperCDMS\cite{Schnee:2005pj}, ZEPLIN-MAX\cite{Atac:2005wv}, and LUX\cite{lux} which are
expected to reach a sensitivity of $10^{-45}$ cm$^2$ or more.
Finally, we note that several of the patterns are well separated in the $\sigma(\tilde\chi_1^0 p)$- LSP mass
plots,  providing important signatures along with the  signatures from colliders
for mapping out the sparticle  parameter space.

{\it Acknowledgements:} This work is supported in part by NSF grant  PHY-0456568.


\end{document}